\documentclass[review]{elsarticle}

\usepackage{lineno,hyperref}
\usepackage{amsmath}
\usepackage{amssymb}
\usepackage{amsthm}
\usepackage{yhmath}
\usepackage{booktabs} 
\usepackage{xspace}
\usepackage{graphicx,psfrag,epsf}
\usepackage{float}
\usepackage{caption}
\usepackage{subcaption} 
\graphicspath{ {./figures/} }
\usepackage{xcolor} 
\usepackage{hyperref}
\hypersetup{colorlinks,
citecolor=blue
}
\usepackage{geometry}
\makeatletter
\def\ps@pprintTitle{%
  \let\@oddhead\@empty
  \let\@evenhead\@empty
  \let\@oddfoot\@empty
  \let\@evenfoot\@oddfoot
}
\makeatother

\modulolinenumbers[5]

\biboptions{authoryear}

\newcommand{\packagename}{\texttt{SpatialGEV} \citep{chen.etal21}}
\newcommand{\SI}{\href{https://github.com/meixichen/SI_for_fast_inference_spatial_GEV}{Supplementary Material}}
\newcommand{\appendixrefdelta}{Appendix~\ref{app:delta-method}}
\newcommand{\appendixrefpred}{Appendix~\ref{app:prediction}}
\newcommand{\appendixrefdiff}{Appendix~\ref{app:sim-estimated-errors}}

\newcommand{\hmc}{HMC-NUTS\xspace}
\newcommand{\lapmq}{Laplace-MQ\xspace}
\newcommand{\lapmc}{Laplace-MCMC\xspace}
\newcommand{\msmq}{MaxSmooth-MQ\xspace}
\newcommand{\msmc}{MaxSmooth-MCMC\xspace}
\DeclareMathOperator{\argmin}{arg\,min}
\DeclareMathOperator{\argmax}{arg\,max}
\newcommand{\bm}[1]{\boldsymbol{#1}}
\newcommand{\tx}[1]{\textrm{#1}}

\DeclareMathOperator{\cov}{Cov}
\DeclareMathOperator{\mae}{MAE}
\DeclareMathOperator{\ks}{KS}
\newcommand{\ind}{\overset{\tx{ind}}{\sim}}
\newcommand{\xx}{\boldsymbol{x}}
\newcommand{\yy}{\boldsymbol{y}}

\newcommand{\XX}{\boldsymbol{X}}

\newcommand{\tth}{\boldsymbol{\theta}}

\newcommand{\uu}{\boldsymbol{u}}
\newcommand{\eeta}{\boldsymbol{\eta}}

\newcommand{\bbe}{\boldsymbol{\beta}}
\newcommand{\SSi}{\boldsymbol{\Sigma}}
\newcommand{\VV}{\boldsymbol{V}}

\newcommand{\JJ}{\boldsymbol{J}}

\makeatletter
\newcommand{\ud}{\mathop{}\!{\operator@font{d}}}
\makeatother

\theoremstyle{definition}
\newtheorem*{remark*}{Remark}

\begin{document}

\begin{frontmatter}

\title{Fast and Scalable Inference for Spatial Extreme Value Models}

\author[mainaddress]{Meixi Chen}

\author[mainaddress]{Reza Ramezan}

\author[mainaddress]{Martin Lysy\corref{mycorrespondingauthor}}
\cortext[mycorrespondingauthor]{Corresponding author}
\ead{mlysy@uwaterloo.ca}

\address[mainaddress]{Department of Statistics and Actuarial Science, University of Waterloo, Canada}

\begin{abstract}
The generalized extreme value (GEV) distribution is a popular model for analyzing and forecasting extreme weather data. To increase prediction accuracy, spatial information is often pooled via a latent Gaussian process (GP) on the GEV parameters. Inference for GEV-GP models is typically carried out using Markov chain Monte Carlo (MCMC) methods, or using approximate inference methods such as the integrated nested Laplace approximation (INLA). However, MCMC becomes prohibitively slow as the number of spatial locations increases, whereas INLA is only applicable in practice to a limited subset of GEV-GP models. In this paper, we revisit the original Laplace approximation for fitting spatial GEV models. In combination with a popular sparsity-inducing spatial covariance approximation technique, we show through simulations that our approach accurately estimates the Bayesian predictive distribution of extreme weather events, is scalable to several thousand spatial locations, and is several orders of magnitude faster than MCMC. A case study in forecasting extreme snowfall across Canada is presented.
\end{abstract}

\begin{keyword}
Generalized extreme value distribution\sep latent spatial Gaussian process\sep Laplace approximation \sep sparse precision matrix \sep Bayesian inference.
\end{keyword}

\end{frontmatter}

\section{Introduction}\label{sec:intro}

Statistical modelling of extreme weather data has important applications including efficient wind energy management and damage prevention for floods and hurricanes. Extreme value theory (EVT) provides the tools for analyzing such data, covering a wide range of applications from climatology \citep{casson99, fawcett06, blanchet11} to insurance and finance \citep{embrechts97, gilli06}. Developed under the EVT assumptions, the generalized extreme value (GEV) distribution is commonly used for analyzing a sequence of maxima within non-overlapping time periods. Hence, the GEV distribution is also known as the block maxima model. \cite{coles01} provided an overview of the properties and applications of the GEV distribution with an emphasis on the analysis of hydrological and meteorological data. 

Weather extremes are distinguished from other types of extremes due to the potential presence of spatial patterns. To incorporate the additional spatial information such as longitude and latitude into the analysis, Bayesian hierarchical modelling is a natural approach. The work of \cite{smith87} was the first to show the advantages of Bayesian analysis over likelihood-based methods in some cases of extreme value modelling. \cite{coles96} reviewed the early studies of Bayesian models of extreme values, and discussed the use of Bayesian methods to pool spatial information when data are sparse in the spatial domain. Later, \cite{coles98} presented a simulation study of hurricane wind speeds where each GEV parameter was linked to the spatial covariates via a linear combination of a regression function and a Gaussian process. \mbox{\cite{cooley07}} modelled precipitation extremes as conditionally independent given the GEV model parameters, which are assumed to follow spatial Gaussian processes. Adding a copula model in the data layer, \cite{sang-gelfand10} dropped the conditional independence assumption, while retaining the spatial Gaussian process layer for the parameters. \cite{reich-shaby12} and \cite{stephenson-etal16} proposed spatial max-stable models whose marginal distribution is GEV, and thus can be viewed as infinite-dimensional generalizations of the GEV model. \cite{davison12} provide a comprehensive review of Bayesian hierarchical models of spatial extremes, discussing their development in terms of methodology and theoretical framework. Specifically, they compare three classes of models: latent variable models, copulas, and max-stable models. While latent variable models have the flexibility to specify a host of different latent variable structures, they have been criticized for the unrealistic assumption of conditional independence between nearby locations. In contrast, copula and max-stable models tend to capture the dependency between these locations more efficiently. That being said, \cite{davison12} argue that latent variable models are more useful for estimating marginal (location-specific) quantities such as the expected return level, an important metric for extreme weather analyses (see Section~\ref{sec:method-est-return}).  
This paper, therefore, focuses on latent variable GEV models with parameters modelled via spatial Gaussian processes, which we refer to as GEV-GP models.

There have been numerous applications of the GEV-GP framework to modelling weather extremes in different regions and countries \citep[e.g.,][]{dyrrdal15, tye-cooley15, jalbert.etal17, garcia18, jhannesson-etal22, youngman22}. However, inference for these complex models can be computationally challenging. This is because the computational complexity of likelihood evaluations for the latent Gaussian process scales as $\mathcal{O}(n^3)$, where $n$ is the number of sites being studied.  Therefore, computations for large-scale spatial analyses involving numerous sites can quickly become prohibitively expensive.
This is exacerbated by the fact that the $n$ latent variables corresponding to each location cannot be analytically integrated out. Instead, Markov Chain Monte Carlo (MCMC) methods are often used in the context of Bayesian inference to sample from the joint space of latent variables and model parameters. However, MCMC algorithms can take days or even weeks to converge when the number of spatial random effects is moderate to large.

As an alternative to MCMC, in this paper we develop a Bayesian inference method for GEV-GP models based on the Laplace approximation \citep[e.g.,][]{tierney-kadane86,barndorffnielsen.cox89,breslow.lin95,skaug.fournier06}, a technique for converting an intractable integration problem into an optimization problem which is much easier to solve. The Laplace approximation is not new to spatio-temporal modelling.  Indeed, a refinement of it known as the integrated nested Laplace approximation (INLA) \citep{rue09, rue17} is typically more accurate, and has been recently used to model weather extremes~\citep[e.g.][]{opitz18, castro-camillo19, castro-camillo20}. 
However, the computational complexity of INLA increases exponentially with the number of hyperparameters in the model. In practice, this limitation is so severe that the flagship INLA implementation via the \texttt{R-INLA} package~\citep{lindgren-rue15} can only handle multiple random effects if they enter linearly into the likelihood function, which is not the case for GEV-GP models.
In contrast, the original Laplace approximation imposes no such restriction. We use the Laplace approximation here in combination with a sparsity-inducing approximation of certain spatial Gaussian processes as Gaussian Markov random fields \citep{lindgren11}. 
Our simulation studies indicate that the proposed approach accurately approximates Bayesian posterior predictive return levels for the full range of GEV-GP models, for several thousand spatial locations and several orders of magnitude faster than state-of-the-art MCMC.

The remainder of this paper is structured as follows. Section \ref{sec:method} discusses the details of the GEV-GP model, the Bayesian Laplace approximation, and the sparsity-inducing spatial covariance matrix approximation. Section \ref{sec:simulation} presents 
simulation studies comparing our Laplace approximation to both a state-of-the-art MCMC algorithm, and to another approximate inference method for GEV-GP models known as ``Max-and-Smooth''~\citep{hrafnkelsson-etal19}, which is conceptually similar but typically much faster than Laplace.
Section \ref{sec:case-study} presents 
a case study analyzing extreme monthly total snowfall in Canada. In particular, it highlights the importance of having multiple spatial random effects in the GEV model for accurately predicting extreme weather events, a modelling strategy which cannot be achieved in practice using INLA. 
Concluding remarks are offered in Section \ref{sec:discussion}. Efficient implementations of our methods are provided in the \textsf{R}/\textsf{C++} library \packagename.  
All code to reproduce the results and analyses in this paper are provided in the \SI.

\section{Methodology}\label{sec:method}
The GEV-GP is a hierarchical model consisting of a data layer and a spatial random effects layer. Let $\xx_1, \ldots, \xx_n \in \mathbb{R}^2$ denote the geographical coordinates of $n$ locations, and let $y_{ik}$ denote the $k$th extreme value measurement at location $i$, for $k = 1, \ldots, n_i$. The data layer specifies that each observation $y_{ik}$ has a GEV distribution, denoted by $y \sim \operatorname{GEV}(a, b_o, s_o)$, for which the CDF is given by  
\begin{equation}
    F(y\mid a, b_o, s_o) =
    \begin{cases}
        \exp\left\{-\left(1+s_o\frac{y-a}{b_o}\right)^{-\frac{1}{s_o}}\right\} \ \ &s_o\neq 0,\\
        \exp\left\{-\exp\left(-\frac{y-a}{b_o}\right)\right\} \ \ &s_o=0,
    \end{cases}
\label{eqn:gev-distn}
\end{equation}
where $a\in\mathbb{R}$, $b_o>0$, and $s_o\in\mathbb{R}$ are location, scale, and shape parameters, respectively. The support of the GEV distribution depends on the parameter values: $y$ is bounded below by $a-b_o/s_o$ when $s_o>0$, bounded above by $a-b_o/s_o$ when $s_o<0$.
To avoid imposing an upper bound on extreme weather events, we assume that $s_o > 0$.

In order to capture the spatial dependence in the data, we let the GEV parameters $a(\xx)$, $b_o(\xx)$, and $s_o(\xx)$ be location-dependent and model some or all of them as spatially varying random effects.  In the most general setting, we follow \cite{cooley07} and use independent latent Gaussian processes for each GEV parameter $a(\xx)$, $b(\xx) = \log(b_o(\xx))$, and $s(\xx) = \log(s_o(\xx))$.
A Gaussian process
\begin{equation}
    f(\xx)\sim \mathcal{GP}(m(\xx), k(\xx, \xx'))
\end{equation}
is fully characterized by its mean function $m(\xx) = E[f(\xx)]$ and its kernel function $k(\xx, \xx') = \cov( f(\xx), f(\xx') )$, the latter of which captures the strength of the spatial correlation between locations. Here, we give the mean function 
$m(\xx) = \bm{c}(\xx)^T\bbe$ a regression structure, where $\bm{c}(\xx)$ is a (known) vector-valued function of covariates at location $\xx$ and $\bbe$ is the coefficient vector.  For the kernel function, we choose the Matérn kernel \mbox{\citep{handcock-stein93}} given by
\begin{equation}
    k(\xx, \xx'\mid \sigma^2, \kappa, \nu) = \sigma^2 \frac{2^{1-\nu}}{\Gamma(\nu)}(\kappa ||\xx-\xx'||)^\nu K_\nu(\kappa||\xx-\xx'||),
    \label{eqn:exp-kernel}
\end{equation}
where $K_v(\cdot)$ is the modified Bessel function of the second kind \citep{Abramowitz-Stegun72}, $\sigma^2$ is the variance parameter, $1/\kappa>0$ is the range parameter, and $\nu>0$ is the shape parameter that is typically fixed. Throughout this paper, we let $\nu=1$ \citep{lindgren11}. Thus, $\bbe$, $\sigma^2$ and $\kappa$ are hyperparameters of the Gaussian process.

Given the spatial locations, the data are assumed to follow independent GEV distributions each with their own parameters. The complete GEV-GP hierarchical model is thus
\begin{equation}
\begin{aligned}
y_{ik} \mid a(\xx_i), b(\xx_i), s(\xx_i) & \ind \operatorname{GEV}\big( a(\xx_i), \exp( b(\xx_i) ), \exp( s(\xx_i) )\big)\\
a(\xx) \mid \bbe_a, \sigma_a, \kappa_a &\sim \mathcal{GP}\big( \bm{c}_a(\xx)^T\bbe_a, \ k(\xx, \xx' \mid \sigma_a^2, \kappa_a) \big)\\
b(\xx) \mid \bbe_b, \sigma_b, \kappa_b &\sim \mathcal{GP}\big( \bm{c}_b(\xx)^T\bbe_b, \ k(\xx, \xx' \mid \sigma_b^2, \kappa_b) \big)\\
s(\xx) \mid \bbe_s, \sigma_s, \kappa_s &\sim \mathcal{GP}\big( \bm{c}_s(\xx)^T\bbe_s, \ k(\xx, \xx' \mid \sigma_s^2, \kappa_s) \big).
\end{aligned}
\label{eq:fullmodel}
\end{equation}
The hyperparameters of the model are
\[
\tth=(\bbe_a, \log(\sigma_a^2), \log(\kappa_a),\bbe_b, \log(\sigma_b^2), \log(\kappa_b),\bbe_s, \log(\sigma_s^2), \log(\kappa_s)).
\]

\subsection{Scalable Likelihood Evaluations}\label{sec:scalable-likelihood}
Let $\yy_i = (y_{i1}, \ldots, y_{in_i})$  denote the extreme value observations at location $\xx_i$, and let $a_i = a(\xx_i)$, $b_i = b(\xx_i)$, and $s_i = s(\xx_i)$ denote the corresponding random effects. Let $\yy=(\yy_1, \ldots, \yy_n)^T$, $\bm{a}=(a_1, \ldots, a_n)^T$, $\bm{b}=(b_1, \ldots, b_n)^T$, $\bm{s}=(s_1, \ldots, s_n)^T$, and $\uu = (\bm{a}, \bm{b}, \bm{s})$. Then the joint distribution of data and random effects is
\begin{equation}
\begin{split}
    p(\yy, \uu \mid \tth) &= \prod_{i=1}^n\prod_{k=1}^{n_i}\left\{\frac{1}{\exp(b_i)}\left(1+\exp(s_i)w_{ik}\right)^{-\frac{1+\exp(s_i)}{\exp(s_i)}}\exp\left[-\left(1+\exp(s_i)w_{ik}\right)^{-\frac{1}{\exp(s_i)}}\right]\right\}\\
    &\phantom{=\;} \times\frac{1}{\sqrt{(2\pi)^n\mid \bm{\Sigma}_a\mid }}\exp\left[-\frac{1}{2}(\bm{a}-\bm{C}_a\bbe_a)^T\bm{\Sigma}_a^{-1}(\bm{a}-\bm{C}_a\bbe_a)\right]\\
    &\phantom{=\;} \times\frac{1}{\sqrt{(2\pi)^n\mid \bm{\Sigma}_b\mid }}\exp\left[-\frac{1}{2}(\bm{b}-\bm{C}_b\bbe_b)^T\bm{\Sigma}_b^{-1}(\bm{b}-\bm{C}_b\bbe_b)\right]\\
    &\phantom{=\;} \times\frac{1}{\sqrt{(2\pi)^n\mid \bm{\Sigma}_s\mid }}\exp\left[-\frac{1}{2}(\bm{s}-\bm{C}_s\bbe_s)^T\bm{\Sigma}_s^{-1}(\bm{s}-\bm{C}_s\bbe_s)\right],
\end{split}
\label{eqn:complete-ll}
\end{equation} 
where $w_{ik} = (y_{ik}-a(\xx_i))/\exp(b(\xx_i))$, $\bm{C}_r=(\bm{c}_r(\xx_1) \cdots \bm{c}_r(\xx_n))^T$, and $\SSi_r = [k(\xx_i, \xx_j \mid \sigma^2_r, \kappa_r)]_{1\le i,j \le n}$ for $r \in \{a, b, s\}$.

Since the matrix inversions $\SSi_a^{-1}$, $\SSi_b^{-1}$ and $\SSi_s^{-1}$ and corresponding determinants in~\eqref{eqn:complete-ll} are $\mathcal{O}(n^3)$, 
each evaluation of $p(\yy, \uu \mid \tth)$ quickly becomes extremely expensive as the number of spatial locations $n$ increases.  
One approach to this problem \citep[e.g.,][]{sang08,schliep10,cooley10} is to employ a conditional autoregressive (CAR) model \citep{besag74, besag91} for the random effects, for which the underlying assumption of conditional independence leads to computationally efficient factorizations of the GP contributions to \eqref{eqn:complete-ll}.  
However, the CAR model is only defined at a predetermined set of spatial locations.
Alternatively, \cite{lindgren11} developed an efficient approximation to the GP terms in \eqref{eqn:complete-ll} using a stochastic partial differential equation (SPDE) for the Matérn covariance kernel~\citep[see also][]{opitz17, miller-etal20}.
Specifically, a GP $f(\xx)$ with the Matérn kernel defined in~\eqref{eqn:exp-kernel} can be written as the solution to the SPDE
\begin{equation}
    (\kappa^2 - \Delta)^{\alpha/2}f(\xx) = \mathcal{W}(\xx), \ \ \alpha=\nu+n/2,
    \label{eqn: matern-spde}
\end{equation}
where $(\kappa^2 - \Delta)^{\alpha/2}$ is the differential operator with $\Delta=\sum_{i=1}^2 \partial^2/\partial^2 x_i^2$, and $\mathcal{W}(\xx)$ is a Gaussian white noise process. \cite{lindgren11} showed how to construct a finite element representation of the solution to~\eqref{eqn: matern-spde}, 
which in turn allows one to approximate $\bm{f}= (f(\xx_1), \ldots, f(\xx_n))$ as a Gaussian Markov random field (GMRF), for which covariance matrix inversions and determinant calculations are only $\mathcal{O}(n^{3/2})$ \citep{rue-held05}.  
Moreover, the SPDE approximation readily lends itself to making predictions at new spatial locations whereas the CAR approach does not.

Yet another means of reducing the computational burden is to approximate 
$f(\xx)$ via a basis representation, in the framework of generalized additive models (GAMs)~\citep{wood11,wood-etal16}. This approach has been implemented for GEV models in the \textsf{R} packages \texttt{mgcv}~\citep{wood17,wood22} and \texttt{evgam}~\citep{youngman22}. 
While GAMs typically treat the spatially-varying parameter function $f(\xx)$ as deterministic, there exists a basis expansion for the Matérn GP family with fixed range parameter $1/\kappa$, formulated along similar lines as the SPDE approximation above~\citep{lindgren-rue15}. The random effects of the model are thus reduced to the coefficients of the basis expansion. As the number of these is user-specified and independent of the number of spatial locations, basis expansions provide an invaluable means of scaling approximate GEV-GP inference to massive spatial datasets. However, basis function expansions are notoriously less accurate than keeping all $n$ spatial random effects when the underlying spatial process is not smooth and/or exhibits pockets of short-range correlation~\citep[e.g.,][]{wood20,lindgren-etal21}. In this paper, we therefore focus on the SPDE approximation to achieve scalable evaluations of the joint likelihood function \eqref{eqn:complete-ll}.

\subsection{Estimation of Hyperparameters}\label{sec:est-hyperparameters}
We proceed in a Bayesian context by specifying a prior $\pi(\tth)$ on the hyperparameters of the model in \eqref{eq:fullmodel}.  
The choice of priors for hyperparameters in spatial models has been widely studied. For example, \cite{banerjee-etal03} suggested using informative priors on the variance and range parameters. However, such informative priors typically require domain knowledge from an expert. \cite{simpson-etal17} introduced the so-called PC prior, which penalizes complex models by shrinking the range parameter to infinity and the variance to zero. 
In this paper, we employ a combination of uninformative and weakly informative priors (described fully in Sections \ref{sec:simulation} and \ref{sec:case-study}), 
obtaining good results on real and simulated data. 
Since the focus of this paper is on computations, an in-depth study of different priors is omitted. However, we note that the choice of prior does not limit the applicability of the proposed methods.

Given the prior $\pi(\tth)$, Bayesian inference is typically accomplished using MCMC 
to sample from the joint posterior distribution of hyperparameters and random effects,
\begin{equation}
p(\uu,\tth \mid \yy) \propto p(\yy, \uu \mid \tth) \pi(\tth).
\label{eqn: joint-pos}
\end{equation}
However, the mixing time of MCMC algorithms for the posterior distribution~\eqref{eqn: joint-pos} grows quickly as a function of the number of spatial locations $n$.
As an alternative, 
we now present the Laplace approximation to the marginal hyperparameter distribution
\begin{equation}
p(\tth \mid \yy) \propto \mathcal{L}(\tth \mid \yy) \pi(\tth),
\label{eqn:marg-posterior}
\end{equation}
where  
\begin{equation}
\begin{split}
    \mathcal{L}(\tth \mid \yy) &= \int p(\yy, \uu\mid\tth) \ud \uu \\
    &= \int \exp\big\{-G(\uu; \tth)\big\} \ d \uu. \label{eqn:marginal-ll}
\end{split}
\end{equation}
For the GEV-GP model, this integral is intractable, suggesting the use of MCMC on the joint posterior distribution \eqref{eqn: joint-pos}. Instead, the Laplace approximation converts the intractable integral into a tractable optimization problem. 
It does this by noting that for fixed $\tth$, a second order Taylor expansion of $-G(\uu;\tth) = \log p(\yy \mid \tth) + \log p(\uu \mid \yy, \tth)$ about its mode in $\uu$ approximates the conditional distribution of the random effects given both data and hyperparameters as multivariate normal, which in turn produces a tractable approximation to the marginal distribution $p(\yy \mid \tth)$.
Specifically, for given $\tth$ let 
\begin{equation}\label{eqn:argmax-u}
\uu_{\tth} = \argmin_{\uu} G(\uu; \tth),
\end{equation}
and let us approximate $G(\uu; \tth)$ by its second-order Taylor expansion about $\uu_{\tth}$:
\begin{equation}
    G(\uu; \tth) \approx G(\uu_{\tth}; \tth) + \frac{1}{2}(\uu-\uu_{\tth})^T \bm{H}_{\tth} (\uu-\uu_{\tth}),
    \label{eqn:taylor-expansion}
\end{equation}
where 
\begin{equation}\label{eqn:argmax-hess}
\bm{H}_{\tth}=\frac{\partial^2}{\partial \uu\partial\uu^T}G(\uu; \tth)|_{\uu=\uu_{\tth}}:=\frac{\partial^2}{\partial \uu\partial\uu^T}G(\uu_{\tth}; \tth),
\end{equation}
and the first-order term $\frac{\partial}{\partial \uu} G(\uu_{\tth}; \tth)^T(\uu - \uu_{\tth})$ has vanished since the gradient of $G(\uu; \tth)$ equals zero at the mode $\uu_{\tth}$. Let $\phi(\bm{z} \mid \bm{\mu}, \SSi)$ denote the PDF of $\bm{z} \sim \mathcal{N}(\bm{\mu}, \SSi)$. Substituting the Taylor expansion \eqref{eqn:taylor-expansion} into \eqref{eqn:marginal-ll} gives the Laplace approximation to the marginal likelihood:
\begin{equation}
\begin{aligned}
    \widehat{\mathcal{L}}(\tth\mid \yy) &= \int \exp\left\{-G(\uu_{\tth}; \tth) - \frac{1}{2}(\uu-\uu_{\tth})^T \bm{H}_{\tth} (\uu-\uu_{\tth})\right\} \ud \uu\\
    & \propto \exp\left\{-G(\uu_{\tth}; \tth) - \frac{1}{2}\log\vert \bm{H}_{\tth}\vert \right\} \cdot \int \phi(\uu\vert\uu_{\tth}, \bm{H}_{\tth}^{-1}) \ud \uu\\
    &= \exp\left\{-G(\uu_{\tth}; \tth) - \frac{1}{2}\log\mid \bm{H}_{\tth}\vert \right\}.
\end{aligned}
    \label{eqn:approx-mll}
\end{equation}

The Laplace approximation~\eqref{eqn:approx-mll} leads to the marginal posterior approximation
\begin{equation}
\widehat{p}(\tth \mid \yy) \propto \widehat{\mathcal{L}}(\tth \mid \yy) \pi(\tth).
\label{eqn:posterior-propto}
\end{equation}
In contrast to the joint posterior~\eqref{eqn: joint-pos} on both hyperparameters and random effects, $\widehat{p}(\tth \mid \yy)$ can be explored via MCMC targeting the much smaller set of hyperparameters alone. 
The Laplace approximation can also be used to forego MCMC altogether by constructing 
a Normal approximation to the hyperparameter posterior distribution,
\begin{equation}\label{eqn:normal-approx-theta}
\tth \mid \yy \approx \mathcal{N}(\widehat{\tth}, \ \widehat{\VV}_{\tth}),
\end{equation}
where 
$\widehat{\tth} = \argmax_{\tth} \log \widehat{p}(\tth \mid \yy)$ and $\widehat{\VV}_{\tth} = -\left[\frac{\partial^2}{\partial \tth^2}\log\widehat{p}(\widehat{\tth}\mid \yy)\right]^{-1}$.

The approximation error of the Laplace approximation \eqref{eqn:approx-mll} has been investigated by e.g., \cite{shun-mccullagh95,rue09,ogden21}.  Assuming, for simplicity, that there are an equal number of observations $n_i \equiv n_0$ at each spatial location, the Bayesian normal approximation \eqref{eqn:normal-approx-theta} converges to the true posterior as $n_0 \to \infty$ with $n/n_0 \to 0$, where $n$ is the number of spatial locations, which includes the possibility of $n \to \infty$ as well~\citep{rue09}.

\begin{remark*}
Our treatment of the GP mean coefficient hyperparameters $\bbe_a$, $\bbe_b$, and $\bbe_s$ differs from INLA in that -- provided we use a Gaussian prior for these variables as we have done in Section~\ref{sec:simulation} -- we do not group them together with the random effect parameters $\uu$ to integrate them out of the model. Since the computational cost of INLA increases exponentially with the number of hyperparameters, it is advantageous for it to integrate out as many of these as possible. Since the Laplace approximation has no such limitation, we opted to use it only to integrate over the random effects, thereby obtaining an approximate marginal likelihood for all the hyperparameters of the model.  We could also have used Laplace to integrate over the mean coefficient hyperparameters. While a systematic comparison between the two approaches in terms of speed and accuracy is beyond the scope of this paper, our preliminary investigations indicate that their performance is fairly similar. That being said, with a Gaussian prior, the mean coefficient hyperparameters can be analytically integrated out of the random-effects model, thus leading to a more accurate Laplace approximation than either of the aforementioned alternatives. 
However, performing this integration results in a model for which the random-effects precision matrix is no longer sparse, 
such that the computational benefits of the SPDE approximation are destroyed.
\end{remark*}

\subsection{Estimation of Random Effects}\label{sec:est-random-effects}

As noted in Section~\ref{sec:est-hyperparameters}, the Laplace marginal likelihood is obtained by approximating the conditional random-effects distribution $p(\uu \mid \yy, \tth)$ as
\begin{equation}\label{eqn:u-conditional-normal}
    \uu \mid \yy, \tth \approx \mathcal{N}(\uu_{\tth}, \widehat{\VV}_{\uu}(\tth)),
\end{equation}
where $\widehat{\VV}_{\uu}(\tth) = \bm{H}_{\tth}^{-1}$, and $\uu_{\tth}$ and $\bm{H}_{\tth}$ are given by~\eqref{eqn:argmax-u} and~\eqref{eqn:argmax-hess}.  This suggests a two-step sampling scheme for a Laplace approximation to the random effects posterior distribution $p(\uu \mid \yy) = \int p(\uu \mid \yy, \tth) p(\tth \mid \yy) \ud \tth$, by first drawing a hyperparameter $\tth_i \sim \mathcal{N}(\widehat{\tth}, \widehat{\VV}_{\tth})$ from the Normal hyperparameter posterior approximation~\eqref{eqn:normal-approx-theta}, then drawing $\uu_i \sim \mathcal{N}(\uu_{\tth_i}, \widehat{\VV}_{\uu}(\tth_i))$.  However, this procedure is computationally intensive because a separate numerical optimization must be performed for each $\tth_i$ to find the mean and covariance in~\eqref{eqn:u-conditional-normal}.   Alternatively, we propose a much faster method to approximate $p(\uu \mid \yy)$ as follows.

Given the conditional Laplace approximation~\eqref{eqn:u-conditional-normal}, we apply a first-order Taylor expansion of $\uu_{\tth}$ and a zeroth-order Taylor expansion of $\widehat{\VV}_u(\tth)$ about $\tth=\widehat{\tth}$, such that these two functions are approximated by
\begin{align}
\uu_{\tth} &\approx \uu_{\widehat{\tth}} + \JJ_{\uu}(\tth-\widehat{\tth})\label{eqn:first-zero-taylor-uv}, & 
\widehat{\VV}_u(\tth) &\approx \widehat{\VV}_u(\widehat{\tth}),
\end{align}
where the Jacobian is
\begin{equation}\label{eqn:jacobian-u}
\begin{aligned}
    \JJ_{\uu} & = \left.\frac{\partial}{\partial \tth} \uu_{\tth} \right\vert_{\tth=\widehat{\tth}}
    = \left.-\left[\left(\frac{\partial^2}{\partial \uu\partial \uu^T}G(\uu_{\tth}, \tth)\right)^{-1}\left(\frac{\partial^2}{\partial \uu \partial\tth^T}G(\uu_{\tth}, \tth)\right)\right]\right\vert_{\tth=\widehat{\tth}}.
\end{aligned}
\end{equation} 
Substituting the Taylor approximations in~\eqref{eqn:first-zero-taylor-uv} into the Laplace conditional distribution~\eqref{eqn:u-conditional-normal}, and combining it with the Normal hyperparameter marginal approximation~\eqref{eqn:normal-approx-theta}, one obtains the jointly Normal Laplace approximation 
\begin{equation}
\uu, \tth \mid \yy \approx  
\mathcal{N}\left(
\begin{pmatrix}\uu_{\widehat{\tth}}\\
 \widehat{\tth} 
\end{pmatrix}, \
\begin{pmatrix}
\widehat{\VV}_u(\widehat{\tth}) + \JJ_{\uu}\widehat{\VV}_{\tth}\JJ_{\uu}^T & \JJ_{\uu}\widehat{\VV}_{\tth} \\
 \widehat{\VV}_{\tth}\JJ_{\uu}^T & \widehat{\VV}_{\tth}
\end{pmatrix}
 \right).
\label{eqn:joint-posterior}
\end{equation}
which we refer to henceforth as $\widetilde p(\tth, \uu \mid \yy)$.  The marginal random effects posterior $\widetilde p(\uu \mid \yy)$ arising from~\eqref{eqn:joint-posterior} is
\begin{equation}\label{eqn:normal-approx-u}
   \uu \mid \yy \approx \mathcal{N}\big(\uu_{\widehat{\tth}}, \ 
   \widehat{\VV}_u(\widehat{\tth}) + \JJ_{\uu}\widehat{\VV}_{\tth}\JJ_{\uu}^T\big),
\end{equation}
which coincides with the parametric empirical Bayes (PEB) approximation of \cite{kass-steffey89}. In this manner, the computationally intensive two-step sampling scheme above can be replaced by the much faster method of sampling from the jointly Normal approximate posterior of hyperparameters and random effects~\eqref{eqn:joint-posterior}.  

\subsection{Estimation of Return Levels}\label{sec:method-est-return}
An important application of extreme weather modelling is 
to estimate the $p\times 100$\% upper quantile of the extreme value distribution at a given location $\xx$ \citep{coles01}. For a GEV model with parameters depending on $\xx$, 
we denote this quantity by
\begin{equation}\label{eqn:upper-quantile}
\begin{aligned}
z_p(\xx) &= z_p\big(a(\xx), b(\xx), s(\xx)\big)\\ 
&= F^{-1}(1-p\mid a(\xx), \exp(b(\xx)), \exp(s(\xx))) \\
&= a(\xx)-\frac{\exp(b(\xx))}{\exp(s(\xx))}\left\{1-[-\log(1-p)]^{-\exp(s(\xx))}\right\},
\end{aligned}
\end{equation}
where $F^{-1}(\cdot \mid a, b_o, s_o)$ is the quantile function of the GEV distribution~\eqref{eqn:gev-distn}.
With extreme annual rainfalls as an example, $z_p(\xx)$ is interpreted as the value above which the maximum precipitation level in a given year at location $\xx$ occurs with probability $p$. Once such an event occurs, the expected time for the maximum precipitation to return to this level is $1/p$ years. For this reason, $z_p(\xx)$ is also called the \emph{expected return level} for $1/p$ years. It is an important indicator of how extreme an event might be at a given location when $p$ is chosen to be a small number. 

In the context of the GEV-GP model, suppose we now wish to estimate $z_p(\xx_i)$, $i=1,\ldots,n$, for each of the $n$ spatial locations in a given dataset.  
For problems where $n$ is small to moderate, 
approximately sampling from each posterior return level distribution $p(z_p(\xx_i) \mid \yy)$ can be achieved by first sampling from the approximate random effects posterior $\widetilde{p}(\uu \mid \yy)$ in~\eqref{eqn:normal-approx-u}, then applying the transformation~\eqref{eqn:upper-quantile} to each resulting draw of $(a(\xx_i), b(\xx_i), s(\xx_i))$. 
While this strategy works well when $n$ is relatively small, for large $n$, sampling from $\widetilde{p}(\uu \mid \yy)$ requires $n^2$ locations in memory to store the covariance matrix of the marginal posterior in~\eqref{eqn:normal-approx-u}.  
We will briefly discuss how to handle this problem below (Section \ref{sec:implementation}).

\subsection{Implementation}\label{sec:implementation}
The calculation of the posterior mode $\widehat{\tth} = \argmax_{\tth} \log \widehat{p}(\tth \mid \yy)$ of the Laplace marginal approximation~\eqref{eqn:posterior-propto} is a nested optimization problem, with the inner optimization $\uu_{\tth} = \argmin_{\uu} G(\uu; \tth)$ being performed at each step of the outer optimization of $\widehat{p}(\tth \mid \yy)$. 
Moreover, each step of the outer optimization problem requires the calculation of $\log \vert\bm{H}_{\tth}\vert$, where $\bm{H}_{\tth} = \frac{\partial^2}{\partial \uu \partial \uu^T} G(\uu_{\tth}; \tth)$. While, in principle, this can be accomplished with any number of automatic differentiation (AD) programs, careful implementation is required for spatial random effects models due to the large size of $\bm{H}_{\tth}$. Such an implementation is provided by the \textsf{R}/\textsf{C++} library \texttt{TMB} \citep{kristensen16}.  
Not only does \texttt{TMB} use the state-of-the-art sparse Cholesky methods of~\cite{chen.etal08} to calculate $\log \vert\bm{H}_{\tth}\vert$ with low memory overhead, it also provides an efficient computation of $\frac{\partial}{\partial \tth} \log \widehat{p}(\tth \mid \yy)$, 
such that 
$\widehat{\tth}$ can be efficiently obtained using gradient-based algorithms. Furthermore, \texttt{TMB} provides a Delta method calculation (described in \appendixrefdelta) to obtain element-wise posterior means and standard deviations of arbitrary parameter transformations $\bm{\eta}(\tth, \uu) = \big(\eta_1(\tth, \uu), \ldots, \eta_m(\tth, \uu)\big)$, without having to store the $m\times m$ posterior covariance matrix of $\bm{\eta}(\tth, \uu)$ in memory.  This is particularly useful for estimating posterior return levels at all $n$ spatial locations when $n$ is large. 
We provide methods for fitting GEV-GP models via \texttt{TMB} in our \textsf{R}/\textsf{C++} package \packagename.
Our package uses an implementation of the SPDE approximation discussed in Section \ref{sec:scalable-likelihood} provided by the \texttt{R-INLA} package \citep{lindgren-rue15}. 

\section{Simulation Study}\label{sec:simulation} 
In this section, two simulation studies -- consisting of 400 and 6400 spatial locations, respectively -- 
are carried out to demonstrate the speed and accuracy of our Laplace approximation. 
We compare the proposed approximation to two competing Bayesian inference methods. The first is MCMC on the joint posterior distribution of hyperparameters and spatial random effects, of which the covariance matrix is obtained from the SPDE approximation~\eqref{eqn: matern-spde}. We shall refer to this henceforth as the ``true'' posterior distribution. Since this quantity is unavailable in closed form, the accuracy of the Laplace approximation is benchmarked using an MCMC algorithm run until convergence.

Early works on GEV-GP models conduct Bayesian inference using Gibbs sampling on a handful of posterior random variables at a time~\citep[e.g.,][]{cooley07, schliep10, reich-shaby12}, but these methods tend to be inefficient when the latent variables are strongly intercorrelated. In this regard, more efficient block Gibbs samplers have been developed~\citep[e.g.,][]{knorrheld-rue02, geirsson-etal14, geirsson-etal20}. In particular, \cite{jalbert-etal22} note that for GMRFs, conditionally independent spatial locations can be updated in parallel~\citep[e.g.,][]{besag74,jalbert.etal17}, thus leading to substantial computational accelerations when multiple cores with low communication overhead are available. However, these samplers still alternate between updating random effects and hyperparameters, and thus suffer from poor mixing when these quantities are highly correlated. In such situations, joint MCMC updates on all posterior random variables such as Hamiltonian Monte Carlo (HMC)~\citep{duane87} and elliptical slice sampling~\citep{murray10,murray.adams10} typically offer faster convergence. 
In this paper, the MCMC algorithm we employ is the No-U-Turn sampling (NUTS) variant of HMC \citep{hoffman-gelman14} as implemented in the \textsf{R}/\textsf{C++} library \texttt{rstan} \citep{rstan}.  This implementation is heavily optimized for speed and features adaptive hands-free tuning of the NUTS/HMC parameters.  It thus serves to benchmark the Laplace approximation both in terms of accuracy against the true posterior, and in terms of speed compared to a state-of-the-art off-the-shelf MCMC sampler as provided by \texttt{rstan}.

The other Bayesian inference method to which we compare Laplace is the Max-and-Smooth method of~\cite{hrafnkelsson-etal19}. 
Broadly speaking, this method replaces the log-likelihood term of each spatial location by a quadratic approximation -- in the simplest case, using the MLE and Fisher information from only the observations at the given location.  Inference for the random effects model is thus reduced to the much simpler ``Normal-Normal'' model, for which the random effects can be analytically integrated out. 
This principle, seemingly first noted in~\cite{daniels.kass98}, is commonly employed for pooling estimates in meta-analytic studies~\citep[e.g.,][]{sutton-abrams01, nam-etal03, dias-etal18}, and also appears in various applied studies involving hierarchical models~\citep[e.g.][]{daniels.etal00,dominici.etal00,coull.etal03, behseta.etal05,lysy.etal16}.  The Max-and-Smooth methodology provides a unified framework for these approaches with a focus on high-dimensional and complex random-effects structures.

In contrast to the Laplace approximation, Max-and-Smooth performs a single mode-quadrature calculation which can be parallelized across spatial locations, as opposed to making repeated (non-parallelizable) mode-quadrature calculations at each evaluation of the hyperparameter posterior~\eqref{eqn:posterior-propto}. Thus, Max-and-Smooth stands to be much faster than Laplace. It has also been shown to be highly accurate for estimating spatial extreme value model parameters when the number of observations per location is sufficiently large~\citep{hazra-etal21, jhannesson-etal22}.

\subsection{Small-Scale Study on Smooth Surfaces}\label{sec:simulation-medium}

\begin{figure}[!htb]
  \centering
  \begin{subfigure}{.32\textwidth}
	\centering
      \caption{True $a(\xx)$.}\label{fig:sim-maps-true-a}
	\includegraphics[width=\linewidth]{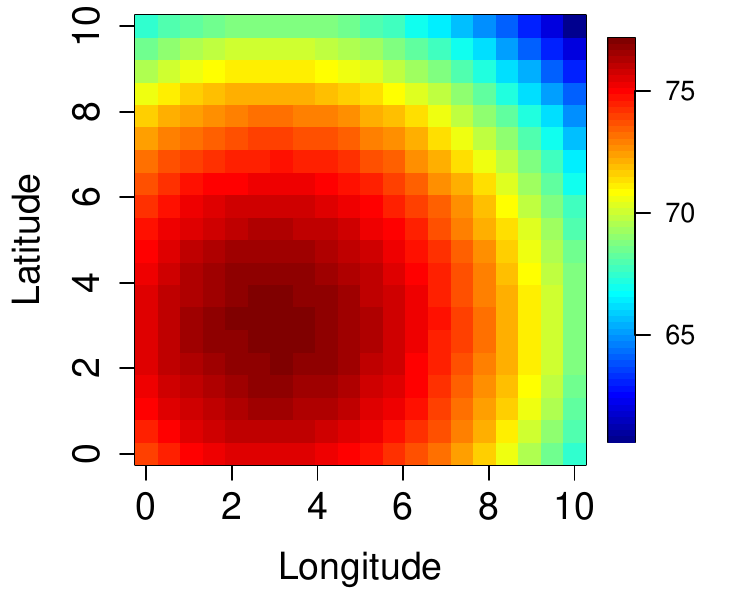}
  \end{subfigure}
  \begin{subfigure}{.32\textwidth}
	\centering
      \caption{True $b(\xx)$.}\label{fig:sim-maps-true-b}
	\includegraphics[width=\linewidth]{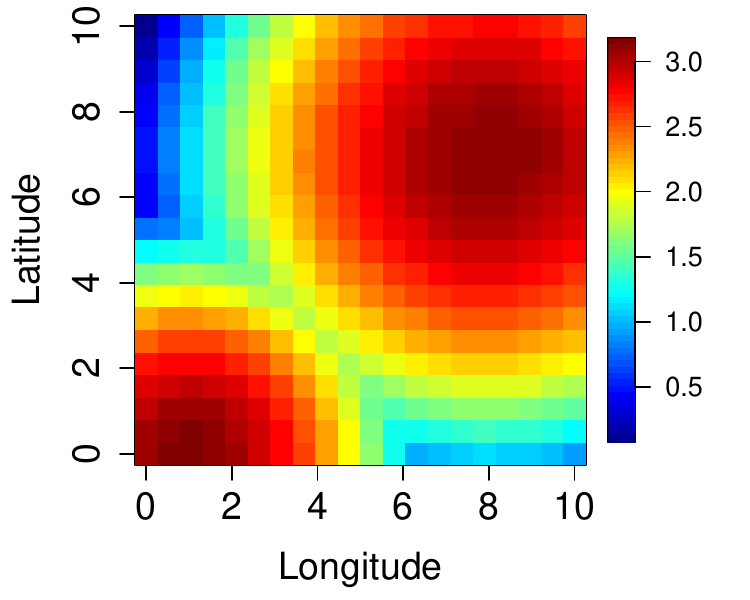}
  \end{subfigure}
  \begin{subfigure}{.32\textwidth}
	\centering
      \caption{True $s(\xx)$.}\label{fig:sim-maps-true-s}
	\includegraphics[width=\linewidth]{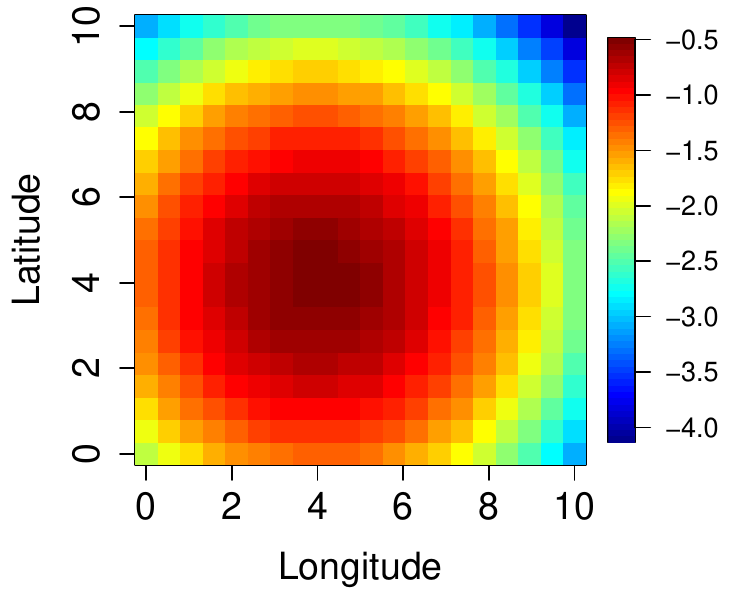}
  \end{subfigure}
  \caption{The true $a(\xx_i)$, $b(\xx_i)$ and $s(\xx_i)$ for the small-scale simulation study plotted on regular lattices.}\label{fig:sim-maps}
\end{figure}

This study consists of spatial locations on a $20\times 20$ regular lattice on $[0, 10] \times [0,10] \subset \mathbb{R}^2$, resulting in a total of $n=400$ locations. The corresponding $a(\xx_i)$, $b(\xx_i)$ and $s(\xx_i)$ are set deterministically via the functions
\begin{align}
a(\xx) = &- 0.5\cdot \log\left\{\tx{det}(2\pi\bm{\Sigma}_0)\right\} - 0.5\cdot (\xx- \bm{\mu}_0)^T\bm{\Sigma}_0^{-1}(\xx- \bm{\mu}_0)+80, \\
\begin{split}
b(\xx) = &0.083\cdot \log\Big\{
0.4\cdot \tx{det}(2\pi\bm{\Sigma}_1)^{-\frac{1}{2}}\exp\left(-\frac{1}{2}(\xx-\bm{\mu}_1)^T\bm{\Sigma}_1^{-1}(\xx-\bm{\mu}_1)\right) \\
            &+ 0.6\cdot \tx{det}(\bm{\Sigma}_2)^{-\frac{1}{2}}
            \exp\left(-\frac{1}{2}(\xx-\bm{\mu}_2)^T\bm{\Sigma}_2^{-1}(\xx-\bm{\Sigma}_2)\right)\Big\}+3.33,
\end{split}\\
s(\xx) = &- 0.1\cdot \log\left\{\tx{det}(2\pi\bm{\Sigma}_3)\right\} - 0.1\cdot (\xx- \bm{\mu}_3)^T\bm{\Sigma}_3^{-1}(\xx- \bm{\mu}_3), 
\end{align}
where 
\begin{equation*}
\begin{aligned}
    \bm{\mu}_0&=\begin{pmatrix}3\\ 3\end{pmatrix}, \ \bm{\Sigma}_0=\begin{pmatrix}3 & 0 \\ 0& 3 \end{pmatrix}, \ \bm{\mu}_1=\begin{pmatrix}1\\ 0\end{pmatrix}, \ \bm{\Sigma}_1=\begin{pmatrix}0.5 & 0 \\ 0& 0.5 \end{pmatrix}, \\ 
    \bm{\mu}_2&=\begin{pmatrix}8\\ 7\end{pmatrix}, \ \bm{\Sigma}_2=\begin{pmatrix}1 & 0 \\ 0& 1 \end{pmatrix}, \ \bm{\mu}_3=\begin{pmatrix}4\\ 4\end{pmatrix}, \ \bm{\Sigma}_3=\begin{pmatrix}2 & 0 \\ 0& 2 \end{pmatrix}.
\end{aligned}
\end{equation*}
Heat maps for $a(\xx)$, $b(\xx)$ and $s(\xx)$ on the spatial domain are displayed in Figure \ref{fig:sim-maps}.
Values of the GEV parameters are chosen so that the simulated observations mimic extreme daily precipitation levels in millimetres \citep{cooley10}. In particular, the range of the shape parameters is $(0.016, 0.6)$, which guarantees that the GEV distribution at each location has a finite mean. At each location $\xx_i$, $n_i$ observations are simulated from $\operatorname{GEV}(a(\xx_i), e^{b(\xx_i)}, e^{s(\xx_i)})$ with $n_i$ drawn from a discrete $\operatorname{Uniform}(20, 50)$. No covariates are included in the mean functions of the spatial Gaussian processes, so each GP has one fixed intercept parameter in the mean function: $\beta_a$, $\beta_b$, and $\beta_s$. We specify weakly informative Normal priors on the GP mean parameters: $\beta_a\sim\mathcal{N}(0, 100^2)$, $\beta_b\sim\mathcal{N}(0,50^2)$, and $\beta_s\sim\mathcal{N}(0, 20^2)$. Flat priors are assumed for 
the hyperparameters of the GP covariance functions. 
We also tried the PC prior of \cite{simpson-etal17} for these but found the difference in parameter estimation to be negligible. 

This simulation study compares five different Bayesian estimators: HMC-NUTS on the joint posterior, and two different estimators for each of Laplace and Max-and-Smooth.  
For Laplace, 
the first of these is the jointly Normal approximation $\widetilde{p}(\tth, \uu \mid \yy)$ in~\eqref{eqn:joint-posterior}, 
which is based on the mode and quadrature of its marginal hyperparameter distribution $\widehat{p}(\tth \mid \yy)$ in~\eqref{eqn:posterior-propto}.   
A jointly Normal mode-quadrature approximation is constructed analogously for Max-and-Smooth.  
For Laplace, we also consider an estimator which samples from its exact posterior distribution $\widehat{p}(\tth, \uu \mid \yy)$, i.e., without the jointly Normal approximation.
To do this, we first sample from 
$\widehat{p}(\tth \mid \yy)$ in~\eqref{eqn:posterior-propto} using MCMC.  Then to each resulting draw of $\tth_i$, we combine a draw $\uu_i$ from the Laplace conditional Normal distribution $\widetilde{p}(\uu \mid \tth, \yy)$ in~\eqref{eqn:u-conditional-normal}, following the two-stage sampling approach described in Section~\ref{sec:est-random-effects}.  Once again, the analogous exact posterior estimator is constructed for Max-and-Smooth.  We shall refer to these five estimators as \hmc, \lapmq, \lapmc, \msmq, and \msmc.  

For the \hmc method, six Markov chains were run in parallel for 4000 iterations per chain, after an initial 4000 ``warm-up'' steps per chain to determine the algorithm's tuning parameters. The effective sample sizes for all fixed and random effects range between $629$ and $38,223$, with a mean of $18,065$. We checked that all chains have mixed well based on the $\widehat{R}<1.01$ convergence metric advocated in \cite{vehtari21}.  

For the Max-and-Smooth methods, following~\cite{jhannesson-etal22}, the Max step at each location was performed using a generalized likelihood estimator~\citep{martins.stedinger00}.  That is, we maximize the log-posterior of the GEV likelihood defined by~\eqref{eqn:gev-distn}, in terms of our restricted parameterization $(a, b = \log b_o, s = \log s_o)$, combined with a prior $s \sim \mathcal{N}(0, 100^2)$ that prevents the mode from being on boundary of the parameter space $s_o = 0$.  

For the \lapmc and \msmc methods, the MCMC algorithms used to explore the marginal hyperparameter posteriors were multivariate Normal random walks with variances proportional to the marginal variance of the corresponding mode-quadrature approximation.  For each of these estimators, six chains were run for 40,000 iterations each, following 5,000 steps of burn-in.

\begin{figure}[!htp]
  \centering
  \includegraphics[width=\linewidth]{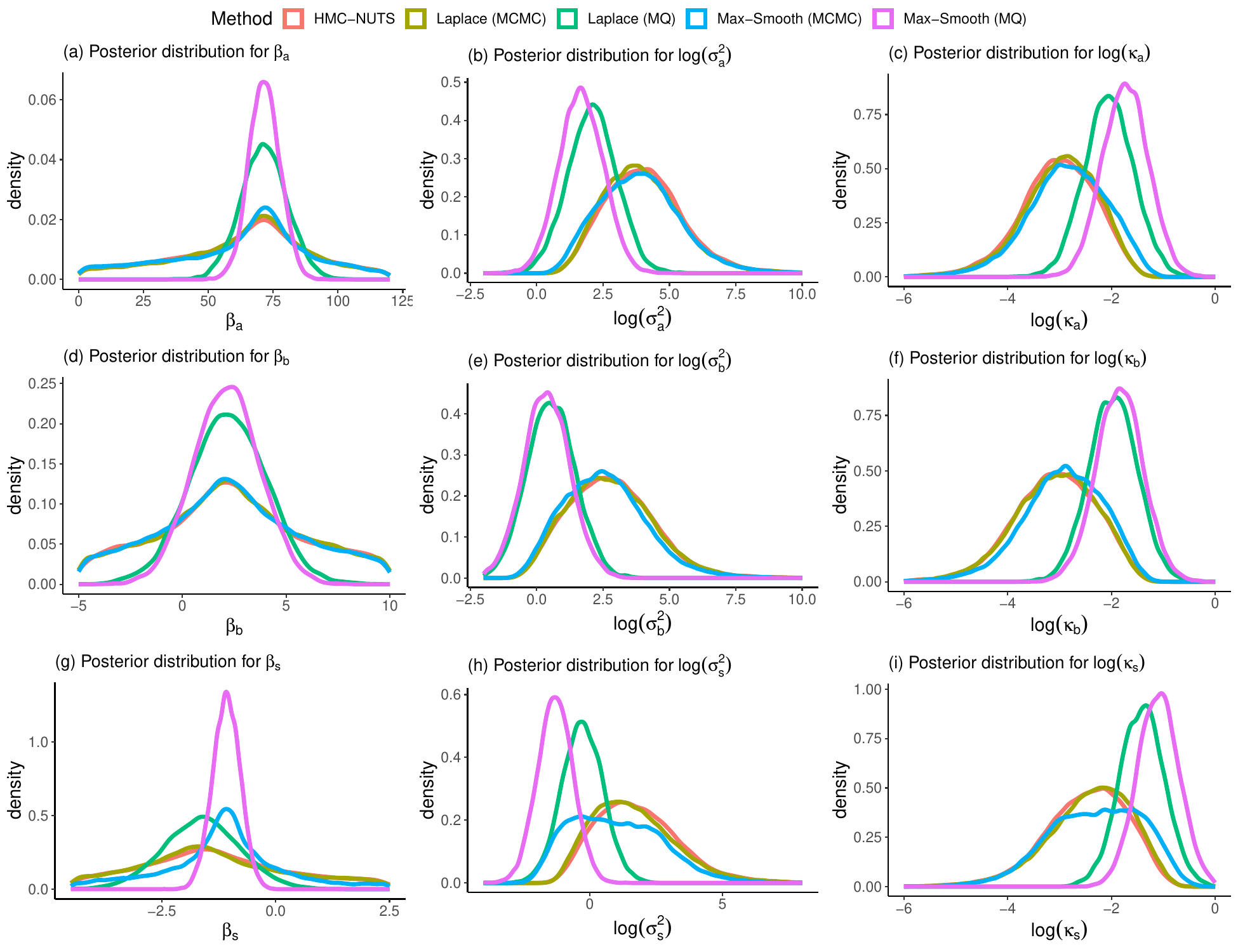}
  \caption{Posterior distributions of the hyperparameters $\tth$ in the small-scale simulation study.}\label{fig:sim-hist-fixed}
\end{figure}

Posterior distributions of the hyperparameters for each of the five Bayesian estimators are shown in Figure~\ref{fig:sim-hist-fixed}.
Figures~\ref{fig:sim-hist-fixed}(a) and~\ref{fig:sim-hist-fixed}(d) show that the two mode-quadrature (MQ) methods \lapmq and \msmq have the same mode of $\beta_a$ and $\beta_b$ as the true posterior obtained from \hmc.  The middle and right columns of Figure \ref{fig:sim-hist-fixed} show that the MQ methods underestimate the means of the variance parameters $(\sigma_a^2, \sigma_b^2, \sigma_s^2)$ and overestimate the means of the inverse range parameters $(\kappa_a, \kappa_b, \kappa_s)$ compared to \hmc.  Moreover, the MQ posteriors for all hyperparameters are much narrower than the corresponding true posteriors.  In contrast, the \lapmc and \msmc estimators are extremely accurate for all hyperparameters except $\beta_s$.

While the \lapmq and \msmq posteriors are generally quite similar, the former is always closer to the true posterior.  The difference between the two is most pronounced for $\beta_s$, the overall mean of the shape parameter, for which \lapmq is markedly superior.  In fact, \lapmq is considerably closer to the true posterior than even \msmc in this case.

\begin{figure}[htp]
    \vspace{-2em}
    \centering
    \includegraphics[width=\linewidth]{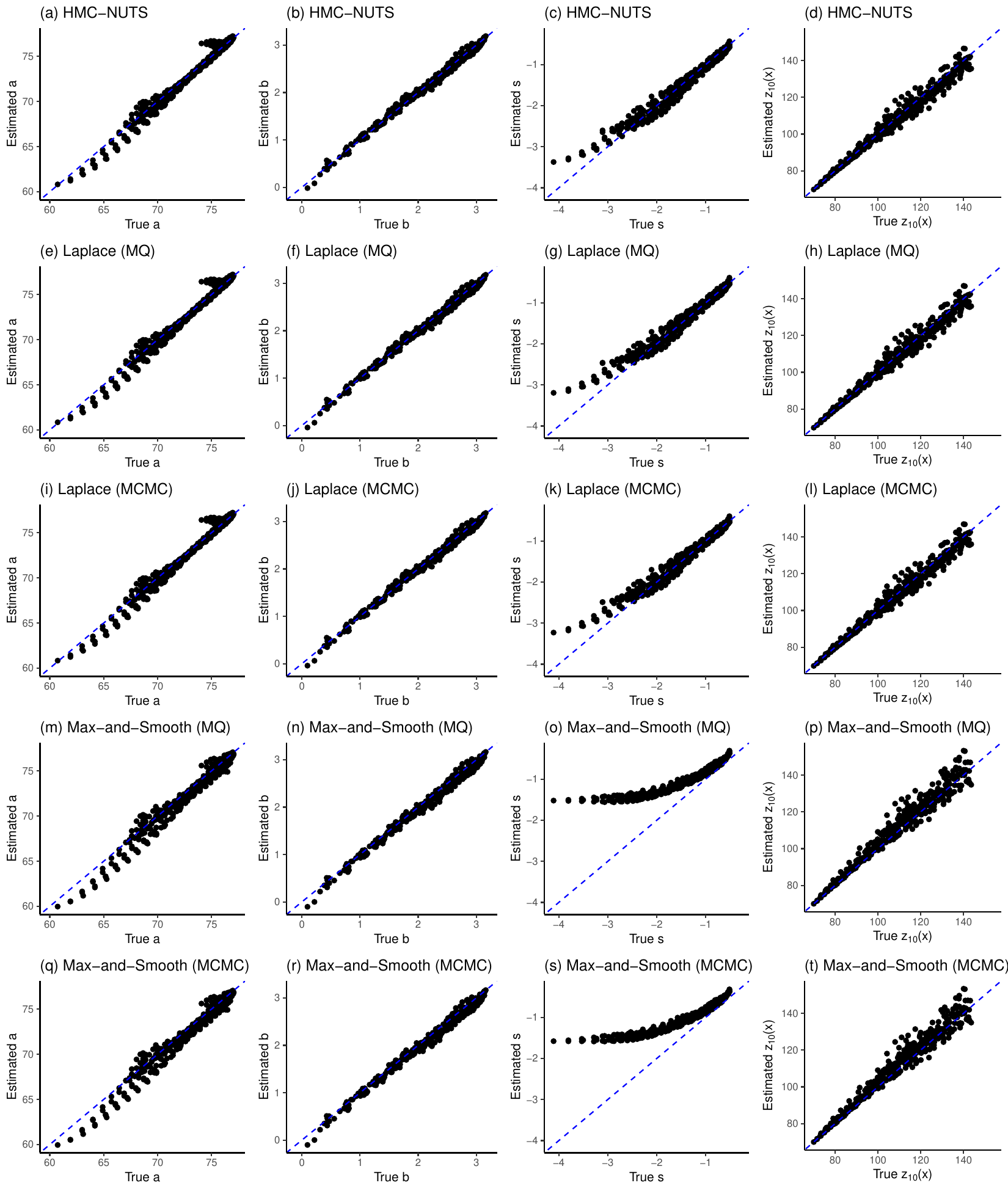}
  \caption{True GEV location parameters $a(\xx)$, scale parameters $b(\xx)$, shape parameters $s(\xx)$, and 10-year return levels $z_{10}(\xx)$ plotted against their estimates obtained using MCMC, Max-and-Smooth, and Laplace.}
  \label{fig:sim-true-vs-est}
\end{figure}

Figure~\ref{fig:sim-true-vs-est} displays the posterior means of the random-effects parameters for each of the five Bayesian estimators against their true values displayed in Figure~\ref{fig:sim-maps}.  
The \lapmq method underestimates the location parameter and correspondingly overestimates the shape parameter at some locations. Further investigations (presented in~\appendixrefdiff) reveal that the locations at which the point estimates of $a(\xx)$ and $s(\xx)$ deviate most from the true values are those lying on the boundary of the spatial domain, and where true boundary values differ substantially from neighbouring values closer to the middle of the domain.  The estimation errors are much worse for the shape parameter than for the location parameter, 
with the former known to be particularly difficult to estimate~\citep[e.g.,][]{schliep10, cooley10}. That being said, the \lapmq posterior mean estimates of the GEV parameters are almost identical to those of the true posterior distribution obtained by \hmc, exhibiting no more than a 1-5\% difference almost invisible in Figure \ref{fig:sim-true-vs-est}.

The \msmq estimates of the location and shape parameters exhibit a similar spatial pattern to those of \lapmq, with the largest error occurring around the boundary and corners of the map.  However, the estimates of the shape parameters are substantially worse than those of \lapmq, and are not much improved with the exact posterior estimator \msmc.


We now consider estimating the return levels $z_p(\xx)$, an important meteorological quantity of interest.  At each of the $n = 400$ spatial locations, the the $10$-year return level $z_{10}(\xx)$ is estimated by its posterior mean, which is computed using the simulation-based method described in 
Section~\ref{sec:method-est-return}.  For the \lapmq and \msmq methods, this is achieved by first drawing $n_\tx{sim}=10,000$ samples from their respective multivariate Normal approximations to $p(\uu \mid \yy)$, then transforming each draw to the corresponding return level via~\eqref{eqn:upper-quantile}.  For the three MCMC estimators, the procedure is identical except using the MCMC draws of $\uu$.

The rightmost column of Figure~\ref{fig:sim-true-vs-est} displays the estimated return levels against their true values.  The \hmc estimator and the two Laplace estimators all exhibit very similar results.  Despite overestimating the shape parameter $s(\xx)$ at its smallest values, the estimates of $z_{10}(\xx)$ by the three aforementioned methods are generally quite accurate.  This suggests that some degree of bias in estimating the shape parameters has only a modest effect on the the marginal quantile estimates when the true shape parameter is close to zero on the original scale.

The return level estimates of the two Max-and-Smooth estimators are also quite accurate, though somewhat less so than the other three estimators at larger return levels.  This is due to the shape parameter being substantially overestimated in the top right corner of Figure~\ref{fig:sim-maps}, which is also where larger values of the scale parameter $b(\xx)$ occur, and thus also of the true return levels $z_{10}(\xx)$.

Table~\ref{tab:comparison} provides a numerical summary of these results, reporting 
the mean absolute errors (MAEs) for $a(\xx)$, $b(\xx)$, $s(\xx)$  and $z_{10}(\xx)$ at the $n=400$ spatial locations. The MAEs for the two Laplace methods differ by no more than 10\% from those of \hmc.  In contrast, the MAEs of \msmq and \msmc are 1.4 to 3.9 times larger than those of \hmc.

While the Laplace and Max-and-Smooth approximations have the same order of error in terms of the number of observations per spatial location~\citep{daniels.kass98}, it is reasonable to expect that the former is more accurate.  This is because the Laplace approximation employs a mode-quadrature approximation to $p(\uu \mid \tth, \yy)$ individually calibrated to each value of $\tth$, instead of a single mode-quadrature approximation calibrated to only the data-level likelihood.  Indeed, Figure 3 in the Supplementary Material shows that for the shape parameter $s$, the standard deviations in the Max step were 1-2 orders of magnitude larger than the parameter estimates themselves in about 10\% of the spatial locations, thus minimizing the contribution of these locations to parameter inference.  This was initially thought to be due to the extremely weak prior $s \sim \mathcal{N}(0, 100^2)$ employed in the generalized likelihood, which is much weaker than the prior on the shape parameter employed by \cite{jhannesson-etal22}.  
Thus, the exact Max-and-Smooth implementation of~\cite{jhannesson-etal22} is applied to this simulation study in the Supplementary Material (Section 5).  It too gives large standard errors on $s$ for many locations, exhibiting lower accuracy than Laplace much like the Max-and-Smooth estimators presented here.  These findings suggest that Max-and-Smooth requires larger sample sizes than Laplace to achieve comparable accuracy.

\begin{figure}[!htp]
  \centering
  \begin{subfigure}{.24\textwidth}
	\centering
    \caption{Posterior SD of $a_{\tx{La}}(\xx)$}\label{fig:sim-sd-a-tmb}
	\includegraphics[width=\linewidth]{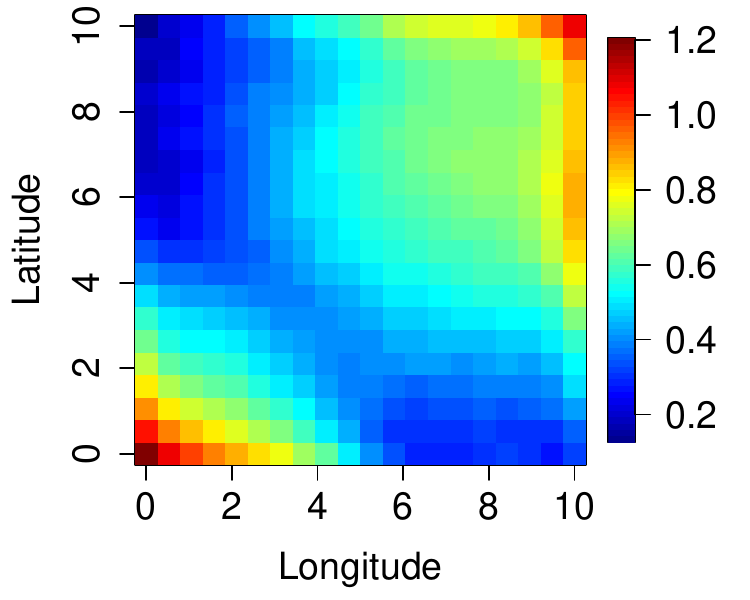}
  \end{subfigure}
  \begin{subfigure}{.24\textwidth}
	\centering
    \caption{Posterior SD of $b_{\tx{La}}(\xx)$}\label{fig:sim-sd-b-tmb}
	\includegraphics[width=\linewidth]{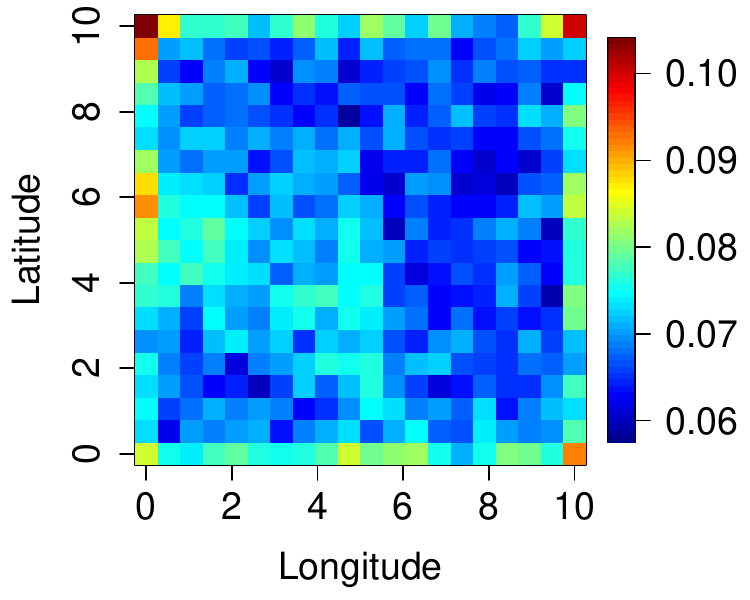}
  \end{subfigure}
  \begin{subfigure}{.24\textwidth}
	\centering
    \caption{Posterior SD of $s_{\tx{La}}(\xx)$}\label{fig:sim-sd-s-tmb}
	\includegraphics[width=\linewidth]{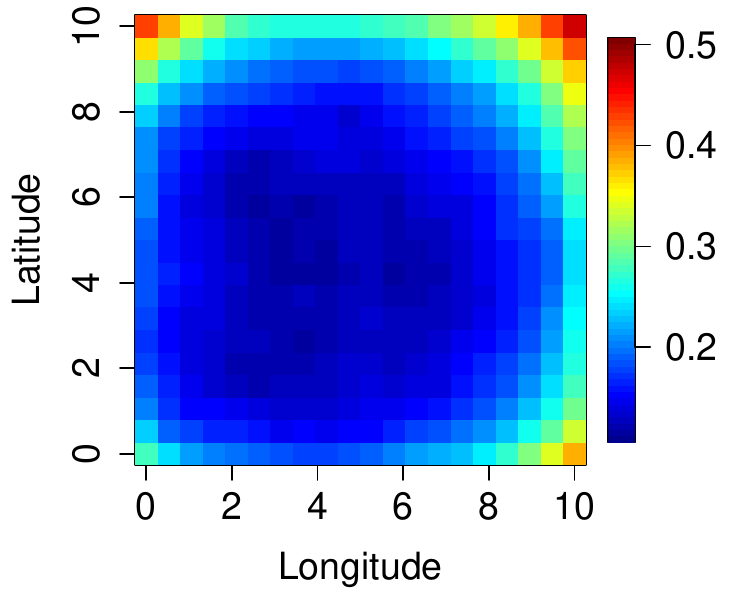}
  \end{subfigure}
    \begin{subfigure}{.24\textwidth}
	\centering
    \caption{Posterior SD of $z_{\tx{La}}(\xx)$}\label{fig:sim-sd-z-tmb}
	\includegraphics[width=\linewidth]{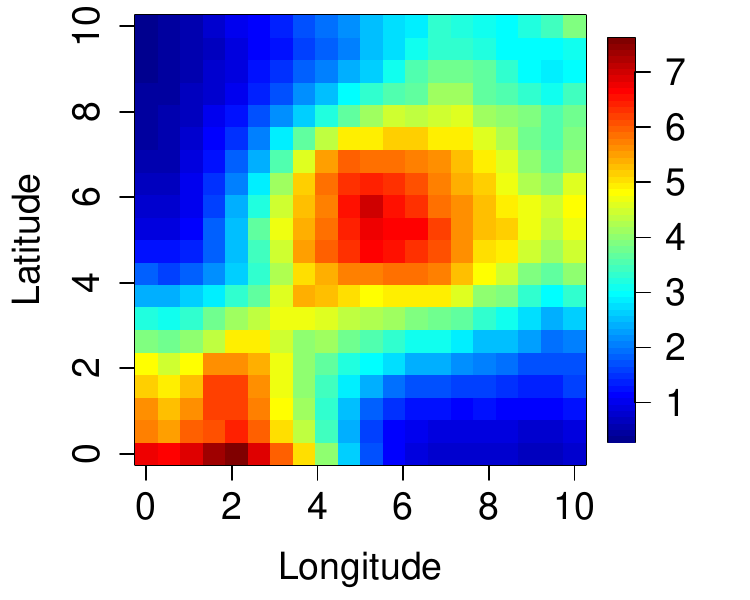}
  \end{subfigure}
  \par\vspace{-1em}
    \begin{subfigure}{.24\textwidth}
	\centering
    \caption{$\frac{\tx{Posterior SD of }a_{\tx{La}}(\xx)} {\tx{Posterior SD of }a_{\tx{MCMC}}(\xx)}$}\label{fig:sim-sd-a-ratio}
	\includegraphics[width=\linewidth]{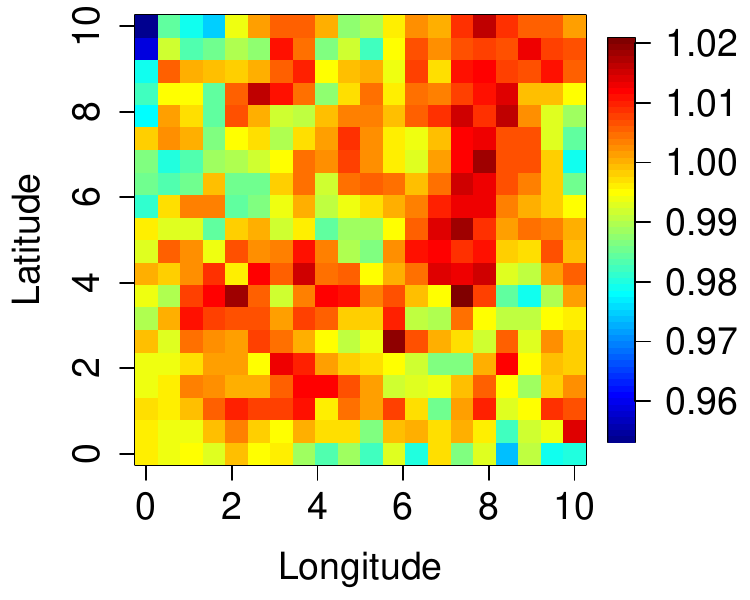}
  \end{subfigure}
  \begin{subfigure}{.24\textwidth}
	\centering
    \caption{$\frac{\tx{Posterior SD of }b_{\tx{La}}(\xx)} {\tx{Posterior SD of }b_{\tx{MCMC}}(\xx)}$}\label{fig:sim-sd-b-ratio}
	\includegraphics[width=\linewidth]{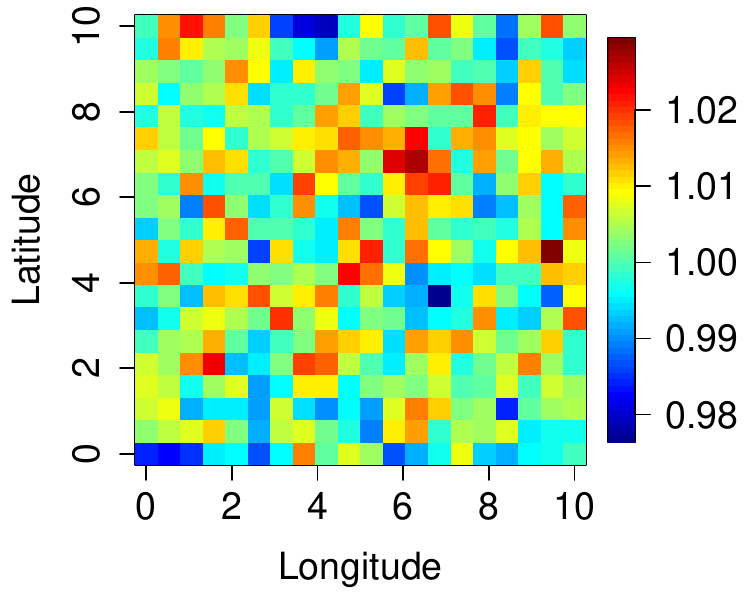}
  \end{subfigure}
  \begin{subfigure}{.24\textwidth}
	\centering
    \caption{$\frac{\tx{Posterior SD of }s_{\tx{La}}(\xx)} {\tx{Posterior SD of }s_{\tx{MCMC}}(\xx)}$}\label{fig:sim-sd-s-ratio}
	\includegraphics[width=\linewidth]{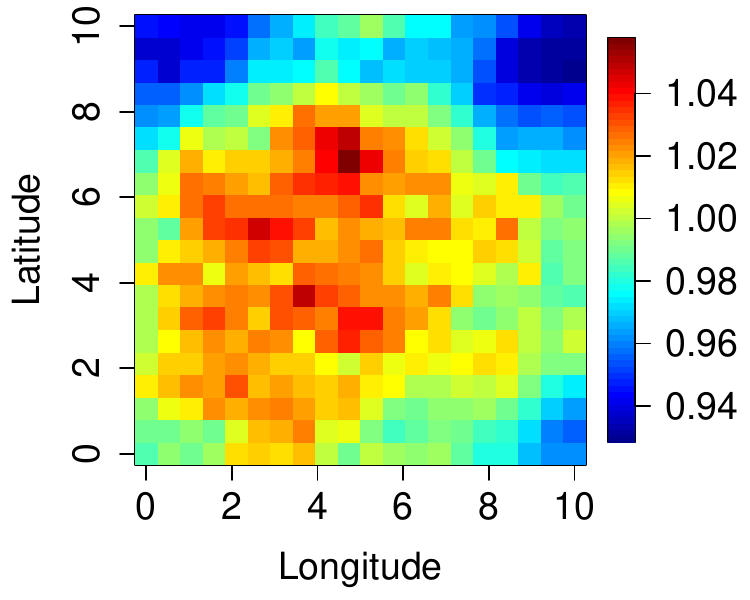}
  \end{subfigure}
  \begin{subfigure}{.24\textwidth}
	\centering
    \caption{$\frac{\tx{Posterior SD of }z_{\tx{La}}(\xx)} {\tx{Posterior SD of }z_{\tx{MCMC}}(\xx)}$}\label{fig:sim-sd-z-ratio}
	\includegraphics[width=\linewidth]{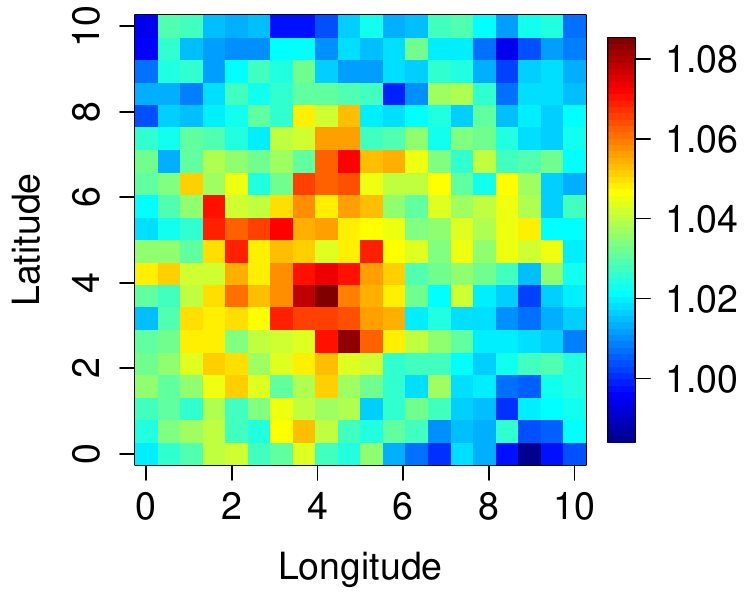}
  \end{subfigure}
  \caption{Upper panels: Standard deviations of posterior samples of $a(\xx_i)$, $b(\xx_i)$, $s(\xx_i)$ and $z_{10}(\xx)$ from the Laplace approach plotted on the regular lattice. Lower panels: Ratios between the posterior standard deviations obtained from the Laplace method and those from MCMC.}\label{fig:sim-SD-maps}
\end{figure}

A well-known shortcoming of the Laplace approximation~\citep[e.g.,][]{kass-steffey89} is that, 
while it typically gives good estimates of the posterior mean of the random effects, it often underestimates their posterior variance. 
To assess this, Figure \ref{fig:sim-SD-maps} examines the uncertainty estimates provided by \lapmq. The top row of Figure \ref{fig:sim-SD-maps} shows the posterior standard deviations of the three GEV parameters at each point on the regular lattice. More uncertainty is observed on the boundary -- especially at the corners -- of the spatial domain.
The bottom row of Figure~\ref{fig:sim-SD-maps} displays the ratio between the posterior standard deviation of \lapmq and that of \hmc. These plots indicate that \lapmq accurately captures the posterior uncertainty in this simulation setting, with all ratios being in the range of 0.95-1.05.

A summary measure of the accuracy of the posterior distributions for $z_{10}(\xx)$ for each of the Laplace nad Max-and-Smooth estimators
is given by the $\ks\big(\widehat{z}_{10}(\xx)\big)$ metric in Table~\ref{tab:comparison}.
This metric is obtained by first computing the two-sample Kolmogorov-Smirnov test statistic (KS) between the posterior draws of $z_{10}(\xx_i)$ from \hmc and those of the given estimator at each location $i = 1,\ldots,400$; then, we average these $n = 400$ test statistics for each estimator to get $\ks\big(\widehat{z}_{10}(\xx)\big)$. 
The more similar are the posterior distributions of $z_{10}(\xx_i)$, the smaller the values of $\ks\big(\widehat{z}_{10}(\xx)\big)$.  By this metric, the Laplace posteriors for $z_{10}(\xx)$ are much more similar to the true posteriors produced by \hmc than those of Max-and-Smooth.

The final metric in Table~\ref{tab:comparison} is the mean computation time for each estimator and its standard error. The latter is calculated from 20 replications of the full estimation procedure for \lapmq and \msmq, and from the six parallel chains run for each of \lapmc, \msmq, and \hmc. All computations were carried out using \texttt{R} version 3.6.2 on a computer cluster with six 2.7GHz Xeon(R) CPUs (E5-4650) and 1GB memory per CPU. The Max step of Max-and-Smooth was parallelized over the six cores, whereas the Smooth step of \msmq, as well as the entirety of the \lapmq calculations, were computed on a single core. 
The runtimes in Table~\ref{tab:comparison} indicate that MCMC on the approximate marginal posteriors of Laplace 
or Max-and-Smooth is considerably faster than on the full joint posterior targeted by \hmc.
\msmc is twice as fast as \lapmc, and would thus be preferable in a setting with enough observations at each spatial location to reliably compute the Max step.  However, both \msmc and \lapmc are orders of magnitude slower than their MQ counterparts.  While the MQ approximations produce biased estimates of the hyperparameters posteriors, \lapmq estimates the random effects and return level posterior distributions with extremely high accuracy.
\begin{table}[!htp]
\centering
\small
\begin{tabular}{@{}rlllll} 
& & \multicolumn{2}{l}{Laplace} & \multicolumn{2}{l}{Max-and-Smooth} \\
\cline{3-6}
& \multicolumn{1}{l}{\hmc} & \multicolumn{1}{l}{MQ} & \multicolumn{1}{l}{MCMC} & \multicolumn{1}{l}{MQ} & \multicolumn{1}{l}{MCMC} \\ 
\hline
$\mae\big(\widehat{a}(\xx)\big)$ & 0.384 & 0.384 & 0.383 & 0.603 & 0.602 \\
$\mae\big(\widehat{b}(\xx)\big)$ & 0.048 & 0.051 & 0.050 & 0.076 & 0.076\\
$\mae\big(\widehat{s}(\xx)\big)$ & 0.102& 0.111 & 0.107 & 0.487 & 0.483\\
$\mae\big(\widehat{z}_{10}(\xx)\big)$ & 2.262 & 2.192 & 2.186 & 3.136 & 3.067\\
$\ks\big(\widehat{z}_{10}(\xx)\big)$ & $0$ & 0.031 & 0.027 & 0.341 &0.340\\
Mean (SD) & - & (0.013) & (0.012) & (0.180) & (0.173)\\
\hline
Runtime & 225,091s & 18s & 70,532s & 14s & 36,011s\\
Mean (SD) & (9110s) & (0.2s) & (114s) & (0.1s) & (116s) \\
\hline
\end{tabular}
\caption{Comparison between \hmc, Laplace (MQ/MCMC), and Max-and-Smooth (MQ/MCMC).  The MAE statistic is computed as $\ks\big(\widehat{u}(\xx)\big) = \frac{\sum_{i=1}^n\vert u(\xx_i)-\widehat{u}(\xx_i)\vert }{n}$, where $u \in \{a, b, s\}$. $\ks\big(\widehat{z}_{10}(\xx)\big)$ is the value of the two-sample KS test statistic between the \hmc and each other posterior for $z_{10}(\xx_i)$ averaged over all spatial locations $i = 1,\ldots, 400$.}
\label{tab:comparison}
\end{table}

\subsection{Large-Scale Study on Rough Surfaces}\label{sec:simulation-large}

\begin{figure}[!htb]
  \centering
  \begin{subfigure}{.32\textwidth}
	\centering
      \caption{True $a(\xx)$.}\label{fig:sim-maps-true-patchy}
	\includegraphics[width=\linewidth]{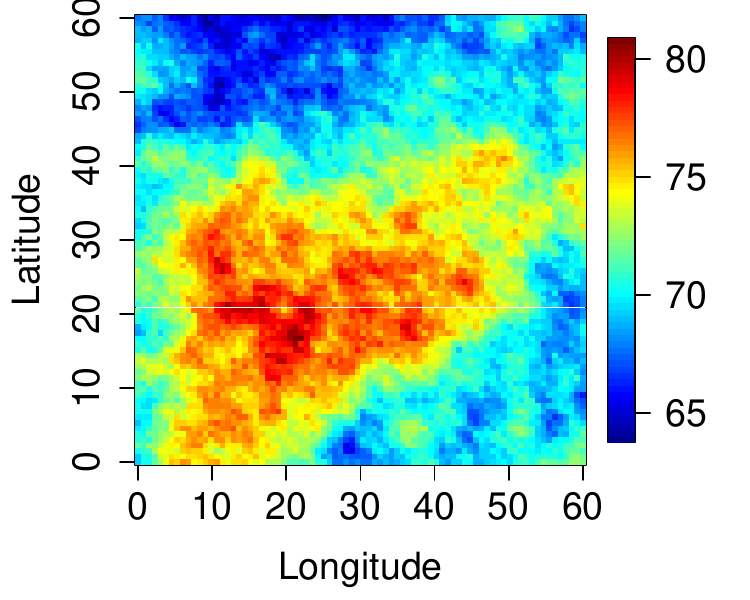}
  \end{subfigure}
  \begin{subfigure}{.32\textwidth}
	\centering
      \caption{True $b(\xx)$.}\label{fig:sim-maps-true-b-patchy}
	\includegraphics[width=\linewidth]{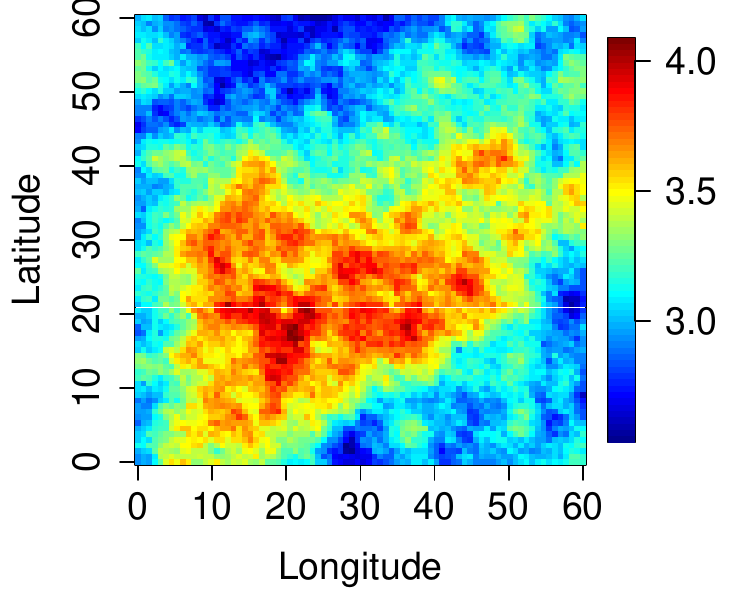}
  \end{subfigure}
  \begin{subfigure}{.32\textwidth}
	\centering
      \caption{True $s(\xx)$.}\label{fig:sim-maps-true-s-patchy}
	\includegraphics[width=\linewidth]{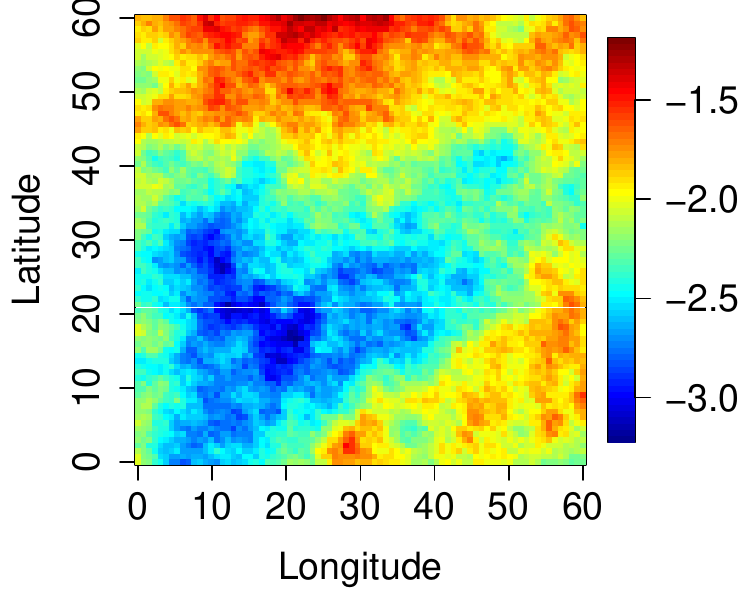}
  \end{subfigure}
  \caption{The true $a(\xx_i)$, $b(\xx_i)$ and $s(\xx_i)$ for the large-scale simulation study simulated from three Gaussian random fields.}\label{fig:sim-maps-patchy}
\end{figure}

The simulation study above has examined the performance of the proposed Laplace method when the GEV parameters are all simulated from relatively smooth surfaces. Here, we consider a more difficult case in which the parameter surfaces are more ``patchy'' (as shown in Figure \ref{fig:sim-maps-patchy}) and where the number of locations is an order of magnitude larger than the previous simulation study. The parameters $a(\xx)$, $b(\xx)$, and $s(\xx)$ are simulated from Gaussian random fields with an exponential covariance kernel function on an $80\times80$ regular lattice on $[0,60]\times[0,60]\subset\mathbb{R}^2$: 
\begin{align}
    a(\xx) &\sim \mathcal{GP}\big(70, \ k(\xx, \xx'\mid 8, 20)\big)\\
    b(\xx) &\sim \mathcal{GP}\big(a(\xx)/10-4, \ k(\xx, \xx'\mid 0.01, 20)\big)\\
    s(\xx) &\sim \mathcal{GP}\big(-a(\xx)/10+5, \ k(\xx, \xx'\mid 0.01, 30)\big),
\end{align} 
where $k(\xx, \xx' \mid \sigma, \lambda)$ is the Mat{\'e}rn covariance kernel~\eqref{eqn:exp-kernel} with shape parameter $\nu = 1$ and scale and range parameters $\sigma$ and $\lambda = 1/\kappa$. Simulation is carried out using the \texttt{rgp()} function in the \texttt{SpatialExtremes} package \citep{ribatet22}. The other aspects of the simulation study -- e.g., data generation mechanism, prior specification, and other modelling setups -- are the same as those in Section \ref{sec:simulation-medium}. 
\begin{figure}[!htb]
    \centering
    \includegraphics[width=\textwidth]{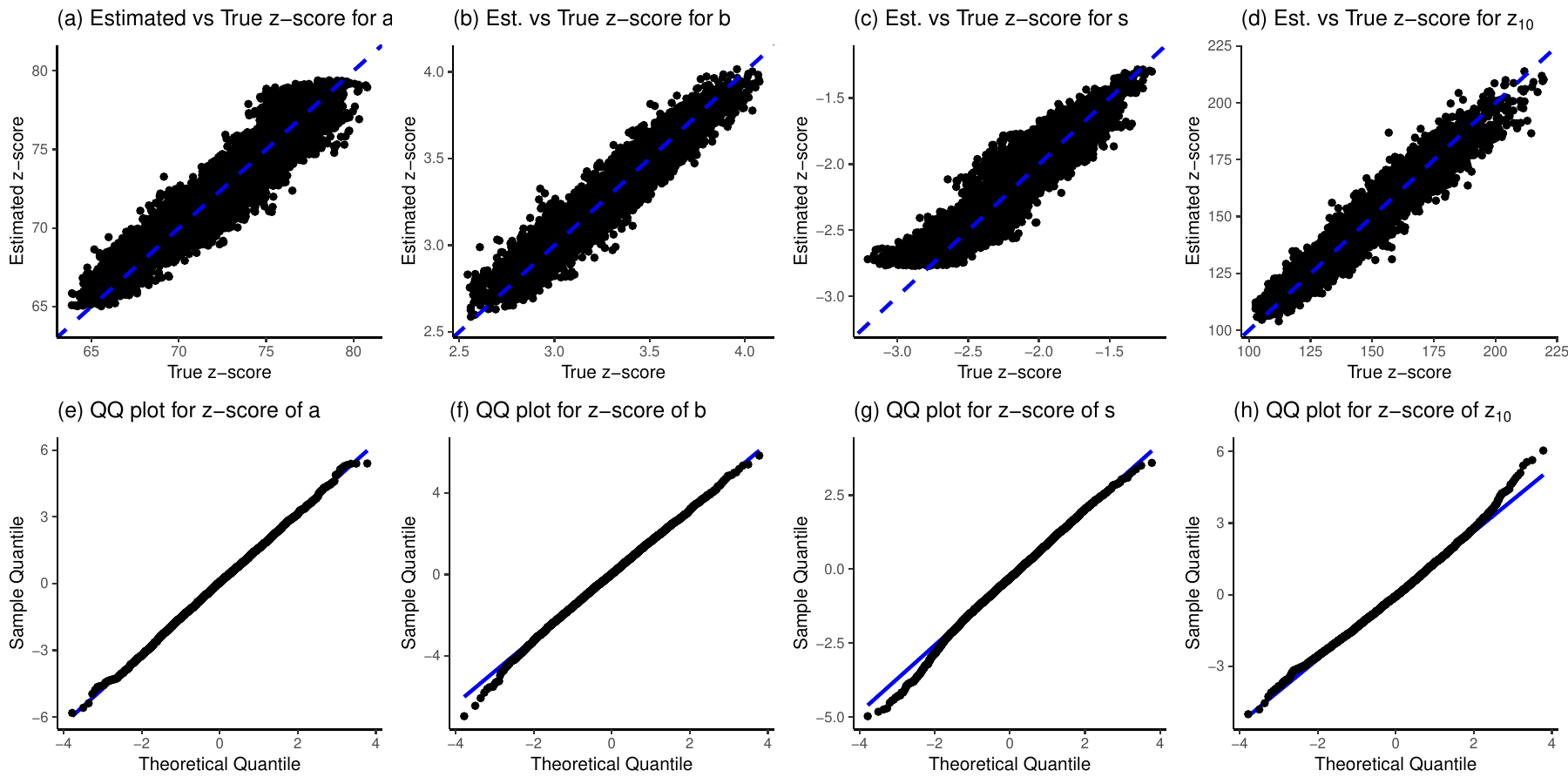}
  \caption{(a)-(d): True GEV location parameters $a(\xx)$, scale parameters $b(\xx)$, shape parameters $s(\xx)$, and 10-year return levels $z_{10}(\xx)$ plotted against their estimates obtained using the Laplace method when the true parameter surfaces are unsmooth. (e)-(f): Q-Q plots for the Z-scores corresponding to $a(\xx)$, $b(\xx)$, $s(\xx)$, and $z_{10}(\xx)$.}
  \label{fig:sim-est-sd-patchy}
\end{figure}

In this simulation, $\uu$ consists of three random effects at each of the $n = 6400$ spatial locations, resulting in a total of $19200$ random effects. Here we only consider the \lapmq estimator, which took $22$ hours to fit on the same computer cluster as used in Section~\ref{sec:implementation} but with 90Gb of memory. As it is computationally infeasible to store and invert the $19200 \times 19200$ dimensional covariance matrix to sample from $p(\uu \mid \yy)$, 
the Delta method is used to compute only the posterior mean and standard deviation of $z_{10}(\xx_i)$ at each spatial location, as discussed in Section \ref{sec:implementation} and \appendixrefdiff. 
As a point of reference, obtaining 100 iterations per parallel chain for \hmc -- not nearly enough to converge to the true posterior distribution -- took over a week. 
While we have not obtained a runtime estimate for \lapmc, the timings in Table~\ref{tab:comparison} suggest it too would be infeasible to run till convergence with the $n = 6400$ locations considered here.

Figure \ref{fig:sim-est-sd-patchy} (a)-(d) compare the posterior mean estimates of all GEV parameters and the 10-year return levels from the Laplace method to their true values. The overall trends of both the location parameter $a(\xx)$ and scale parameter $b(\xx)$ are captured well by the corresponding posterior mean estimates. The shape parameter tends to be slightly overestimated when the true value is small, which has been also observed in Figure \ref{fig:sim-true-vs-est}(c). However, as in the small-scale simulation study, this does not have much of an effect on the estimates of $z_{10}(\xx)$, 
which align well with the true values at the majority of the locations.

We now evaluate the accuracy of the uncertainty estimates of Laplace-MQ.  Since we do not have access to the true posterior distribution for the large-scale simulation, accuracy is assessed by computing at each location $i$ a z-score of the form
\begin{equation}
    \mathcal{Z}(\gamma_i) = \frac{\gamma_i - \hat \gamma_i}{\widehat{\operatorname{sd}}(\gamma_i)},
\end{equation}
where $\gamma_i$ is the true value at location $i$ of the quantity $\gamma \in \{a, b, s, z_{10}\}$, and $\hat \gamma_i$ and $\widehat{\operatorname{sd}}(\gamma_i)$ are the Laplace-MQ estimates of its posterior mean and standard deviation.  If the posterior uncertainty is accurately estimated by Laplace-MQ, then these z-scores should be close to the $45^\circ$ line on a QQ plot.  Such QQ plots are displayed in Figure~\ref{fig:sim-est-sd-patchy} (e)-(h).  The true values of $b$ and $s$ are further than expected in the lower tail of the Laplace-MQ posterior, and those of $z_{10}$ are further than expected in the upper tail.  However, the corresponding departures from the $45^\circ$ line only happen for z-scores of magnitude more than two, meaning that the Laplace-MQ posteriors cover about 95\% of the true underlying values.
These results demonstrate both the computational feasibility of the Laplace-MQ method and its ability to accurately estimate GEV parameters and return levels at several thousand spatial locations.

\section{Case Study}\label{sec:case-study}

Here, we study a real data set of monthly total snowfall in Canada from January 1987 to December 2021, which is publicly available through \cite{snowdata}. The monthly total snowfall at a location is the amount of frozen precipitation (in cm), including snow and ice pellets, observed throughout a month. The goal of this study is to produce a cross-country map of 10-year return levels $z_{10}(\xx)$ of extreme monthly total snowfall at the observed locations.

\begin{figure}[htp!]
    \vspace{-1em}
	\centering
	\begin{subfigure}{.7\textwidth}
	  \centering
	  \caption{All locations}\label{fig:case-study-data-map-raw}
	  \includegraphics[width=\linewidth]{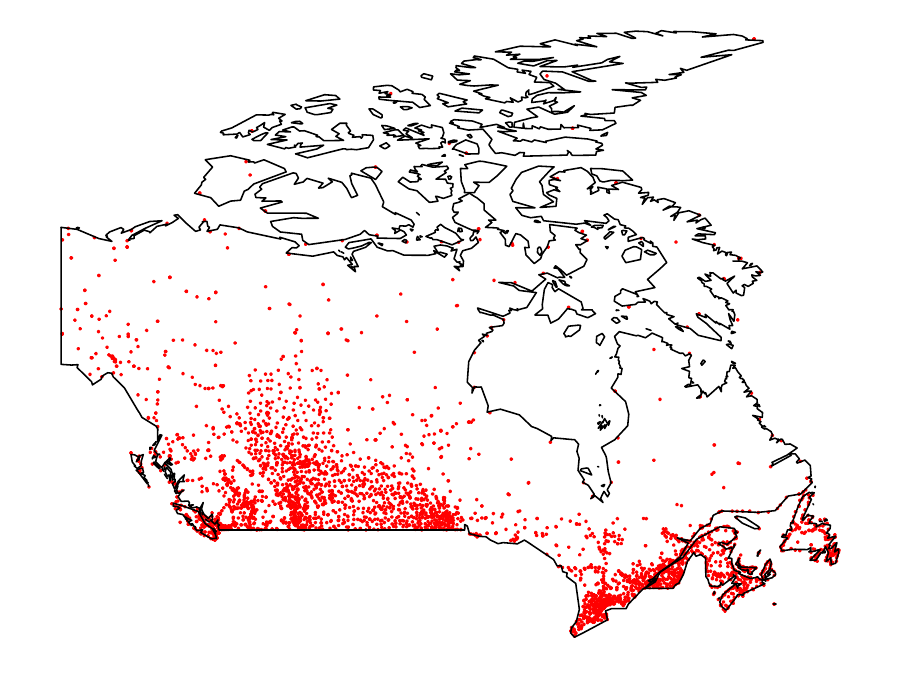}
	\end{subfigure}
	\vspace{-1em}
    \begin{subfigure}{.7\textwidth}
	  \centering
	  \caption{Gridded locations on the mesh}\label{fig:case-study-data-map-grid}
	  \includegraphics[width=\linewidth]{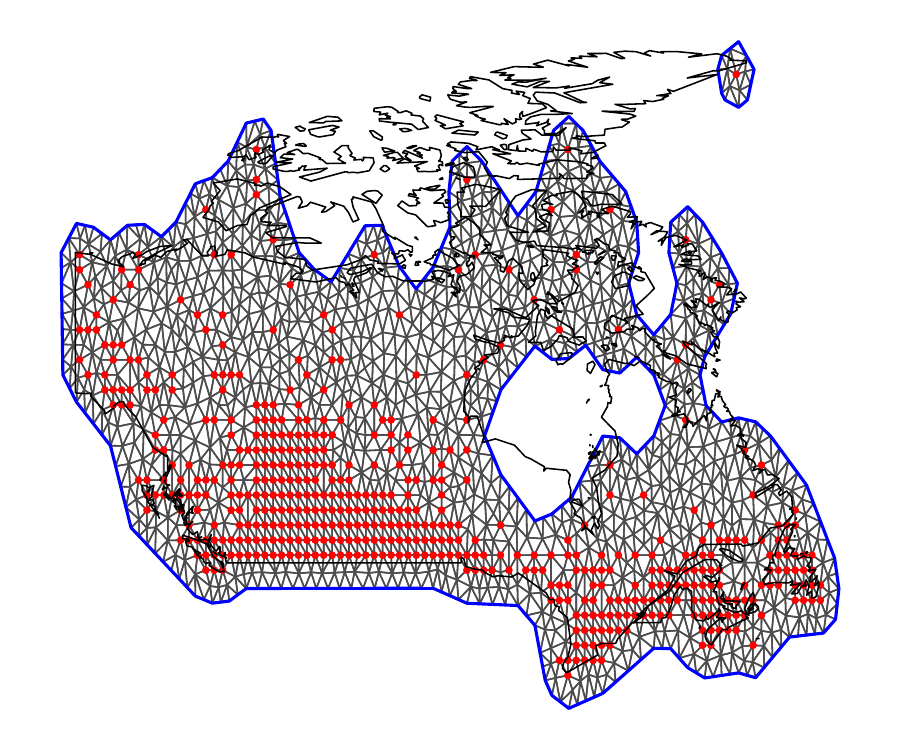}
	\end{subfigure}	
	\caption{(a): All $3833$ locations at which data was collected from 1987 to 2021. (b): $509$ locations at which at least 10 years of data was recorded after gridding. }\label{fig:case-study-data-map}
\end{figure}
The dataset includes $3833$ weather stations plotted in Figure \ref{fig:case-study-data-map}(a). In Figure \ref{fig:case-study-data-map}(b), the data are gridded into $1^{\circ} \times 1^{\circ}$ cells. We only include cells in which there are at least 10 years of observations, and in line with extreme value theory, use the maximum yearly records of monthly snowfall in each cell as the response values. This results $n=509$ spatial locations (grid cells) each with at least 10 extreme value observations. 
{The marginal distribution of these $n=509$ extreme value observations is plotted in Figure~\ref{fig:case-study-obs-hist}. We see that this distribution is heavy-tailed to the right, and thus a good candidate for modelling via the GEV distribution with positive shape parameter.
No additional covariates are included in the model. The triangulated mesh for SPDE approximation is superimposed on the map in Figure \ref{fig:case-study-data-map}(b). 
A non-convex boundary is built for the mesh, leaving out a hole in the region of Hudson Bay due to the physical boundary it creates in the spatial domain. This results in $n'=1565$ nodes, each of which will have separate estimated GEV random effects. Prediction at unobserved locations is straightforward using the Gaussian process methodology \citep{rasmussen-williams06} described in~\appendixrefpred.
\begin{figure}[!htp]
    \vspace{-1em}
    \centering
    \begin{subfigure}{.55\textwidth}
        \centering
        \caption{PDF plot.}
        \includegraphics[width=\linewidth]{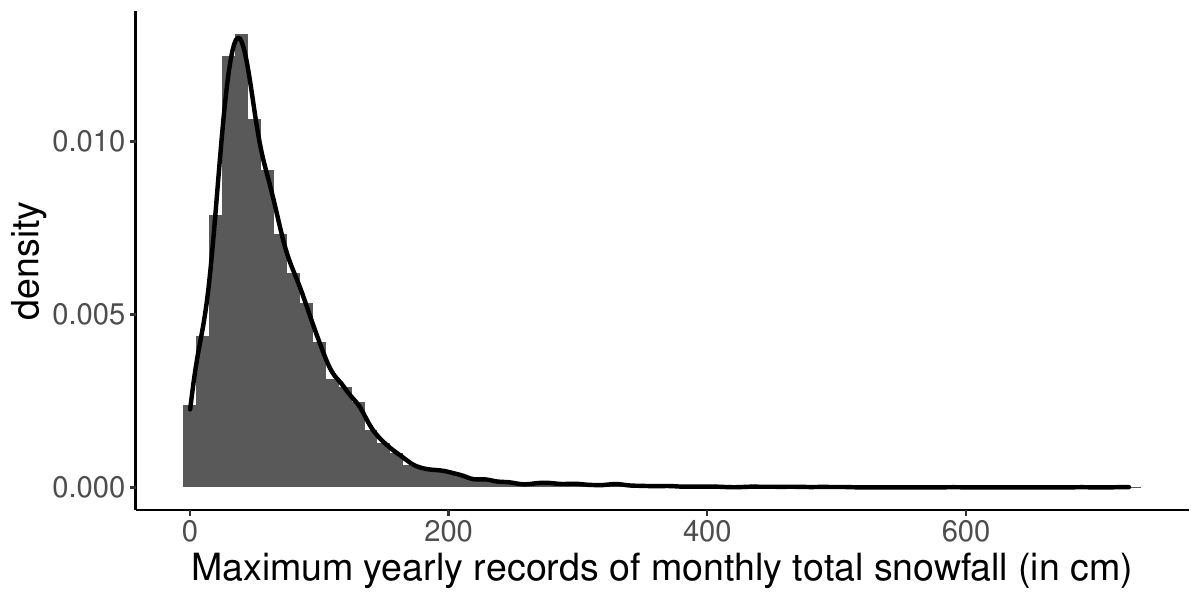}
    \end{subfigure}
    \begin{subfigure}{.44\textwidth}
        \centering
        \caption{CDF plot.}
        \includegraphics[width=\linewidth]{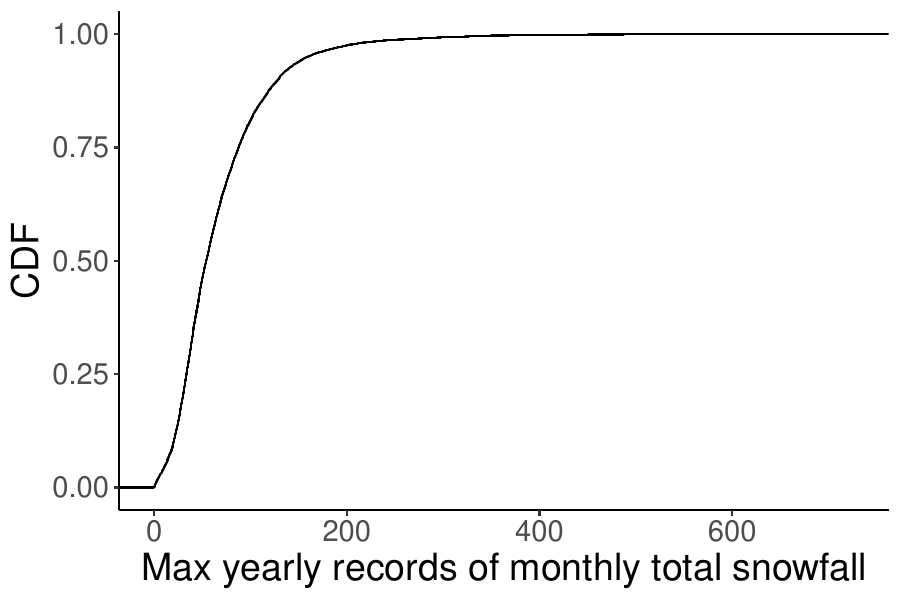}
    \end{subfigure}
    \caption{PDF (left) and CDF (right) of all $n=509$ extreme value observations pooled across locations and time.}\label{fig:case-study-obs-hist}
\end{figure}

\subsection{Model Selection}\label{sec:case-study-select}
The Laplace-MQ method is first used to fit a GEV-GP model with all GEV parameters as spatial random effects, which is denoted by $M_{\tx{abs}}$. The point estimates $\widehat{s}(\xx)$ at all locations have a small range from $-8.1$ to $-8.0$, suggesting that the model with spatially varying $s(\xx)$ might not be needed.
Therefore, we reduce the model complexity by having the shape parameter as a fixed effect $s$ across space, and denote this model by $M_{\tx{ab}}$. A Normal prior $\mathcal{N}(-5,5)$ is imposed on $s$. Furthermore, we compare model $M_{\tx{ab}}$ to the simplest spatial extreme model $M_{\tx{a}}$ with only $a(\xx)$ spatially varying. The model with random scale parameter and fixed location and shape parameters, i.e., $M_{\tx{b}}$, is not considered here as the spatial variation of $a(\xx)$ contributes greatly to the observation variation across different locations.

The fit of models $M_{\tx{ab}}$ and $M_{\tx{a}}$ to the snowfall dataset is evaluated using a model checking procedure described below.
The posterior predictive distribution of maximum monthly total snowfall $y_\star$ at spatial location $\xx_\star$ is given by
\begin{equation}\label{eqn:posterior-pred-distn}
\begin{split}
p(y_\star \mid  \yy) &= \int p\big(y_\star \mid  a(\xx_\star), b(\xx_\star), s(\xx_\star)\big) \times \\
& \ \ \ \ \ \ \ \ \ \ \ \ \ \ \ \ \ p\big(a(\xx_\star), b(\xx_\star), s(\xx_\star) \mid \yy\big) \ud a(\xx_\star) \ud b(\xx_\star) \ud s(\xx_\star).
\end{split}
\end{equation}
Agreement between model and data can be measured by 
computing a chosen test statistic $T_\tx{rep} = T(y_\tx{rep})$, where $y_\tx{rep} \sim p(y_\star \mid \yy)$ are replicated data drawn from the posterior predictive distribution~\eqref{eqn:posterior-pred-distn}, and comparing it to its empirical counterpart $T_\tx{obs} = T(y_\tx{obs})$ computed from the observed data $y_\star = y_\tx{obs}$.
If the model is correctly specified, then $T_{\tx{rep}} \approx T_{\tx{obs}}$. The calculations of the posterior predictive distribution~\eqref{eqn:posterior-pred-distn} and $T_{\tx{rep}}$ are detailed in~\appendixrefpred. 

The test statistics chosen here are the median and the $10$\% upper quantile, i.e. the 10-year return rate. 
Figure \ref{fig:case-study-rep} 
thus plots the median and $10\%$ upper quantiles of the posterior predictive distribution~\eqref{eqn:posterior-pred-distn} ($T_\tx{rep}$) for both models $M_{\tx{ab}}$ and $M_{\tx{a}}$ versus their sample values in the observed data at each of the $n=509$ locations ($T_\tx{obs}$).
Lack of fit is indicated by departures from the $45^{\circ}$ line. 
Reading the plots, it is clear that model $M_{ab}$ outperforms $M_a$.
This observation underscores the importance of having the ability to model more than just the location parameter of the GEV distribution as spatially varying, which cannot be accomplished using existing INLA implementations.

The posterior predictive checks in Figure~\ref{fig:case-study-rep} can also be used to evaluate the goodness-of-fit of the GEV-GP model and its Laplace approximation. We see excellent agreement in the median and upper 10\% quantiles between the actual snowfall values
and the corresponding predictions of model $M_{ab}$, except at the 10-20 locations with the largest snowfall, where it is systematically underestimated.
Consequently, the remaining analyses in this section will be performed with model $M_{\tx{ab}}$.
\begin{figure}[htp!]
  \centering
  \vspace{-2em}
  \begin{subfigure}{0.4\textwidth}
    \caption{Model $M_{\tx{ab}}$: random $a(\xx)$ and $b(\xx)$ \\Test quantity: median.}\label{fig:case-study-rep-50}
    \includegraphics[width=.9\textwidth]{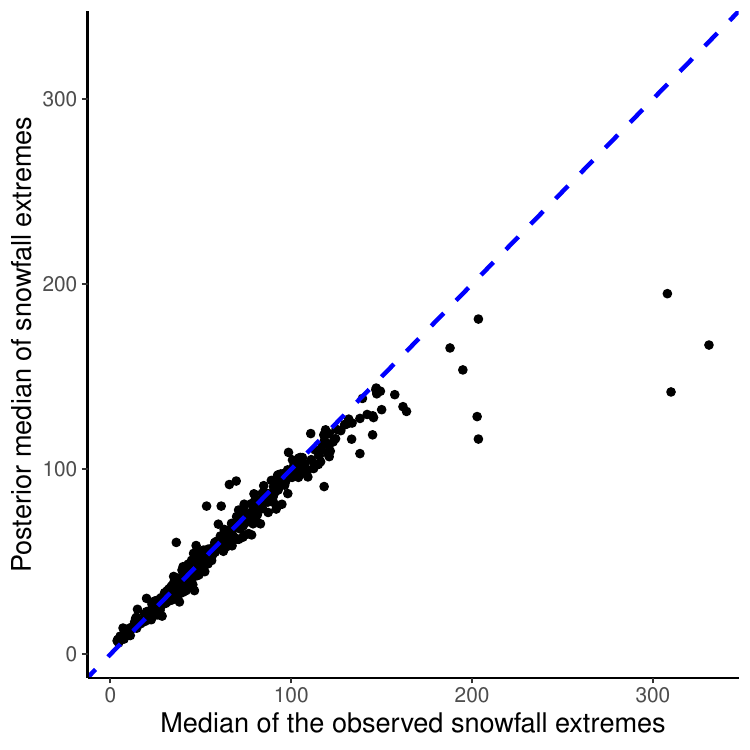}
  \end{subfigure}
  \begin{subfigure}{0.4\textwidth}
    \caption{Model $M_{\tx{ab}}$: random $a(\xx)$ and $b(\xx)$. \\Test quantity: upper $10$\% quantile}\label{fig:case-study-rep-90}
    \includegraphics[width=.9\textwidth]{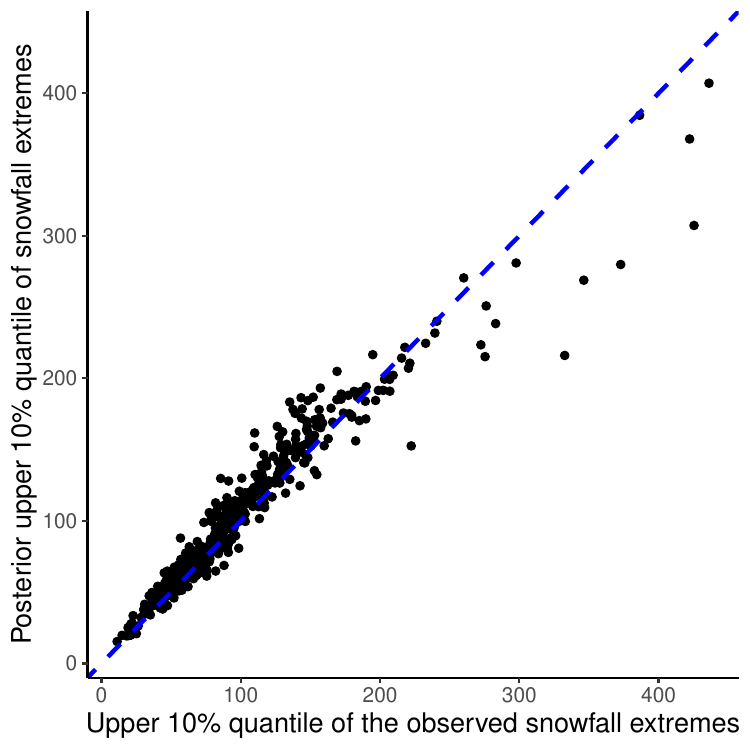}
  \end{subfigure}
  \par\smallskip
  \begin{subfigure}{0.4\textwidth}
    \caption{Model $M_{\tx{a}}$: random $a(\xx)$, fixed $b$. \\Test quantity: median}\label{fig:case-study-rep-50-a}
    \includegraphics[width=.9\textwidth]{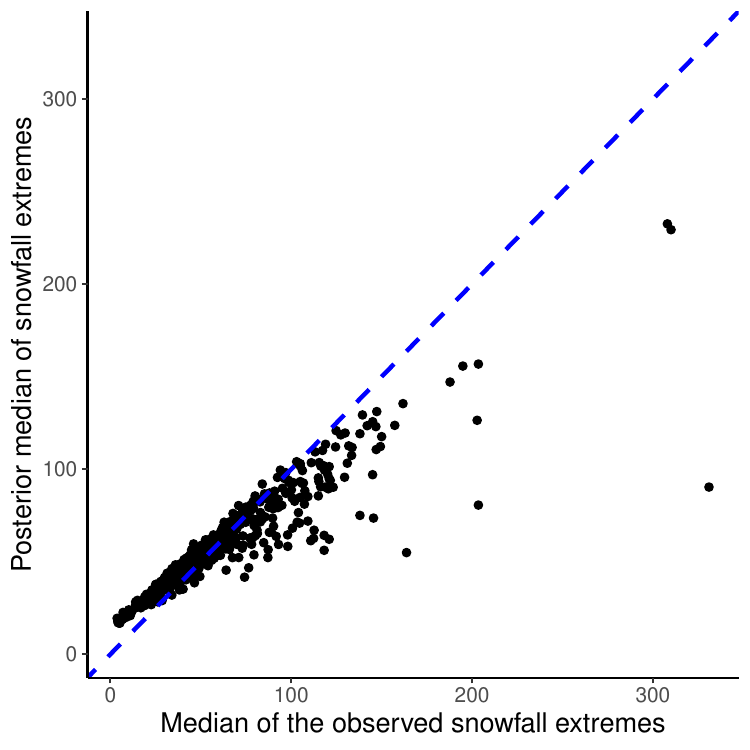}
  \end{subfigure}
  \begin{subfigure}{0.4\textwidth}
    \caption{Model $M_{\tx{a}}$: random $a(\xx)$, fixed $b$. \\Test quantity: upper $10$\% quantile}\label{fig:case-study-rep-90-a}
    \includegraphics[width=.9\textwidth]{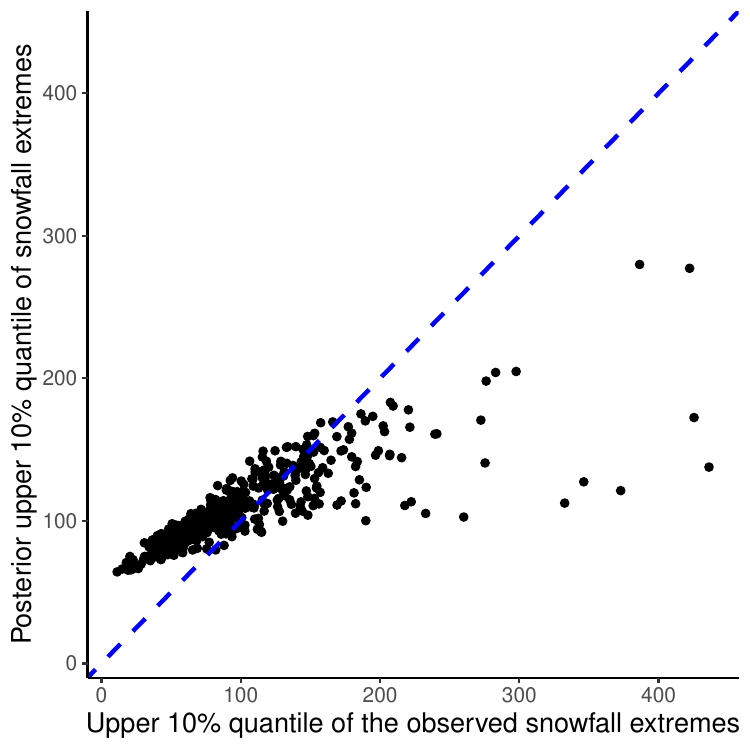}
  \end{subfigure}
  \caption{
  Posterior predicted versus empirical test statistics. Each dot corresponds to one of the $n=509$ spatial locations. Upper panel: Model with both $a(\xx)$ and $b(\xx)$ as spatial random effects. Lower panel: Model with only $a(\xx)$ random. Left panel: Predicted medians versus sample medians. Right panel: Predictive upper $10$\% quantiles versus sample upper $10$\% quantiles.}\label{fig:case-study-rep}
\end{figure}

\subsection{Parameter Inference}

The posterior mean estimate of $s$ is $-7.7$ with a posterior standard deviation of $1.54$. 
While there is some evidence in the simulation studies that the Laplace approximation tends to overestimate $s$ at small values (Figures~\ref{fig:sim-true-vs-est}(g) and~\ref{fig:sim-maps-true-s-patchy}), this only suggests that the true posterior estimate of $s$ could be even smaller. Hence, it would appear that the conditional (i.e., spatially-varying) GEV distribution fitted to these data is in fact light-tailed, thus essentially reducing to a Gumbel distribution~\citep[e.g.,][]{youngman22}. This stands in contrast to the heavy-tailed marginal distribution of the extreme value observations shown in Figure~\ref{fig:case-study-obs-hist}.

\begin{figure}[htp!]
  \centering
  \vspace{-1em}
  \begin{subfigure}{.4\textwidth}
	\centering
	\caption{Posterior mean of $a(\xx)$.}
	\includegraphics[width=\linewidth]{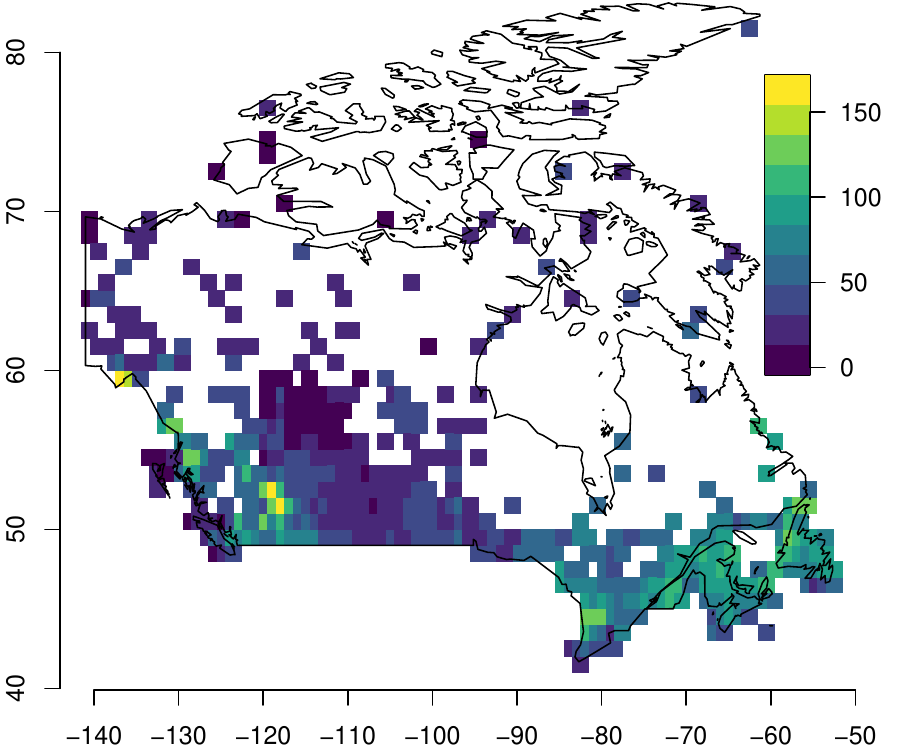}
	\label{fig:case-study-pos-maps-mean-a}
  \end{subfigure}
   \vspace{-1em}
  \begin{subfigure}{.4\textwidth}
	\centering
	\caption{Posterior standard deviation of $a(\xx)$.}
	\includegraphics[width=\linewidth]{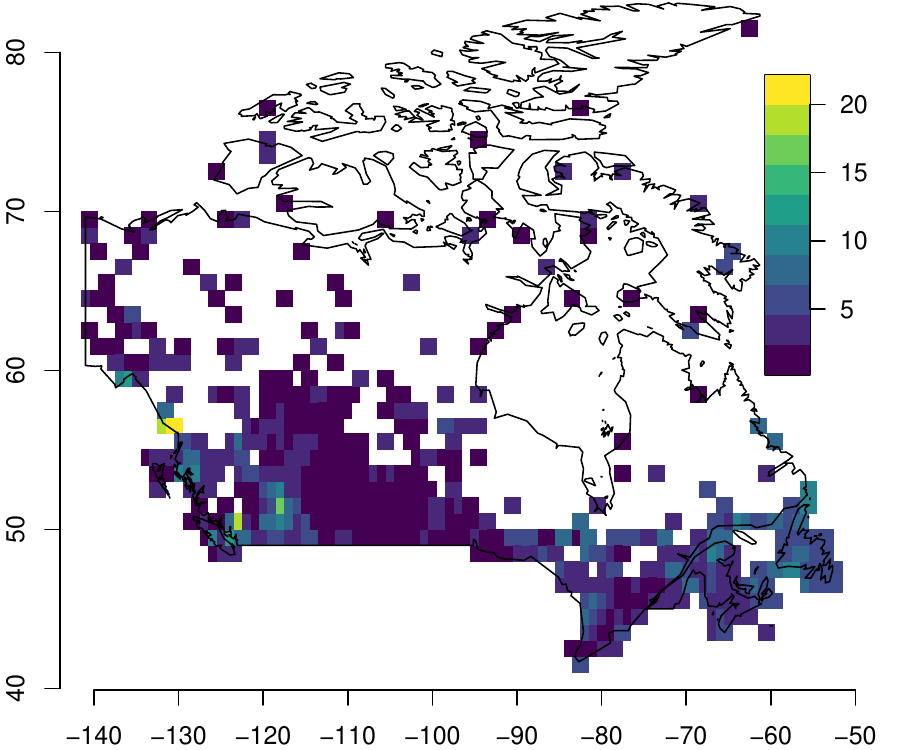}
	\label{fig:case-study-pos-maps-sd-a}
  \end{subfigure}
  \begin{subfigure}{.4\textwidth}
	\centering
	\caption{Posterior mean of $b(\xx)$.}
	\includegraphics[width=\linewidth]{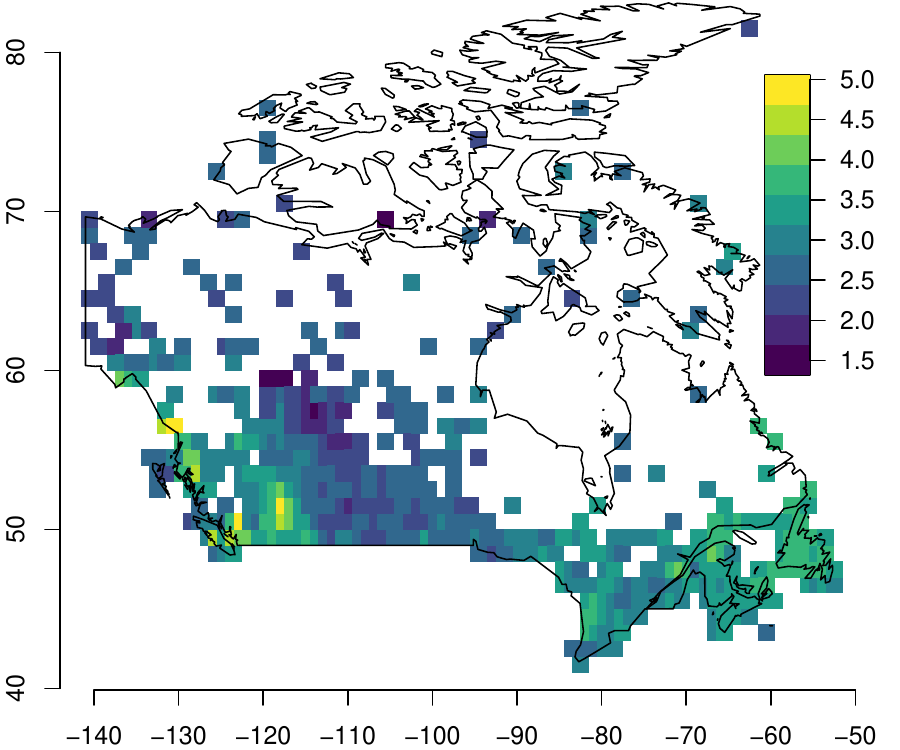}
	\label{fig:case-study-pos-maps-mean-b}
  \end{subfigure}
  \vspace{-1em}
  \begin{subfigure}{.4\textwidth}
	\centering
	\caption{Posterior standard deviation of $b(\xx)$.}
	\includegraphics[width=\linewidth]{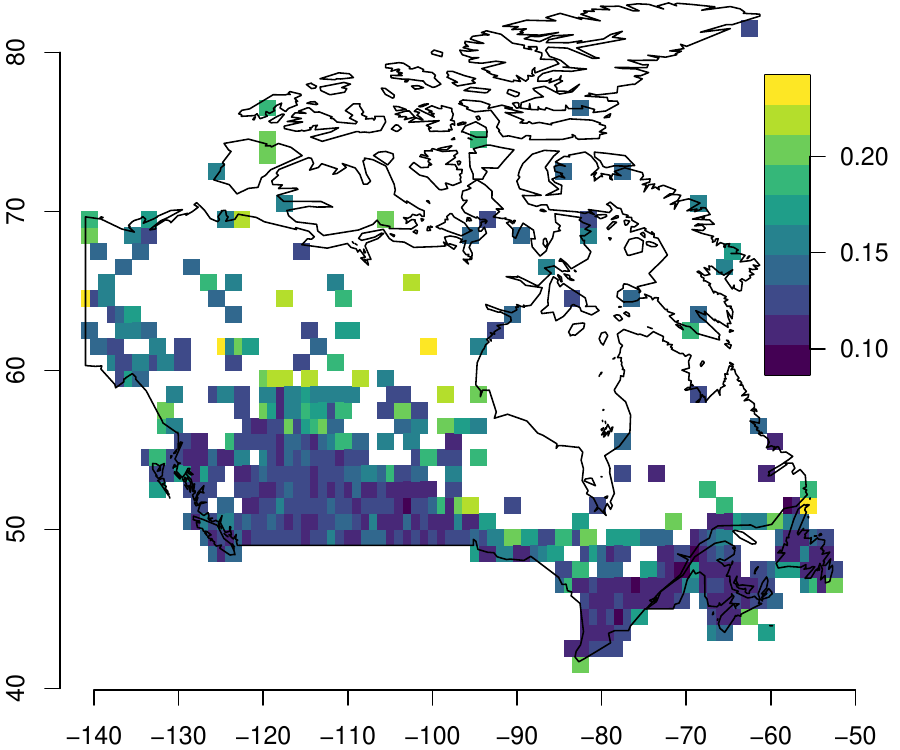}
	\label{fig:case-study-pos-maps-sd-b}
  \end{subfigure}
  \begin{subfigure}{.4\textwidth}
	\centering
	\caption{Posterior mean of 10-year return levels $z_{10}(\xx)$.}
	\includegraphics[width=\linewidth]{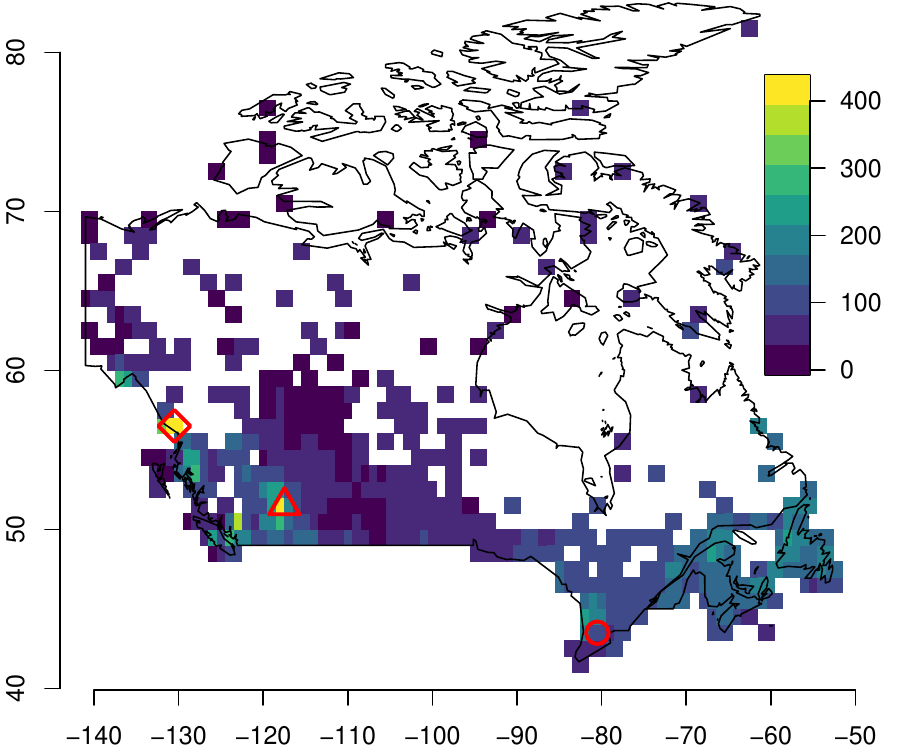}
	\label{fig:case-study-pos-maps-mean-q}
  \end{subfigure}
  \vspace{-1em}
  \begin{subfigure}{.4\textwidth}
	\centering
	\caption{Posterior standard deviation of $z_{10}(\xx)$.}
	\includegraphics[width=\linewidth]{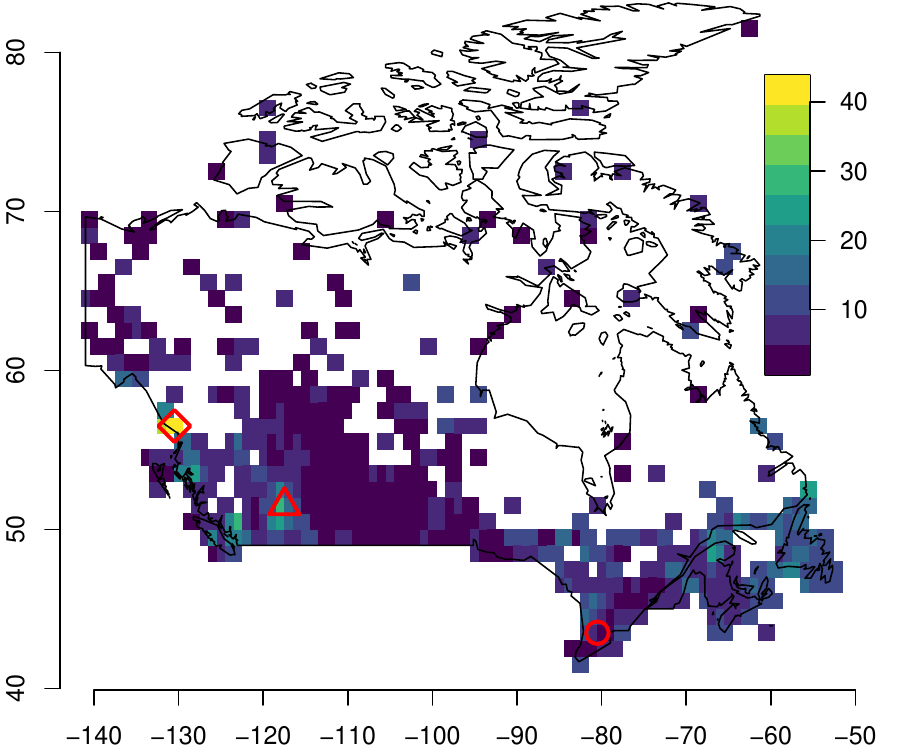}
	\label{fig:case-study-pos-maps-sd-q}
  \end{subfigure}
  \caption{Posterior means and standard deviations of $a(\xx)$, $b(\xx)$, and $z_{10}(\xx)$.}\label{fig:case-study-pos-maps}
\end{figure}
The posterior means and standard deviations of model parameters at each location are plotted in Figure \ref{fig:case-study-pos-maps-mean-a}, \ref{fig:case-study-pos-maps-sd-a}, \ref{fig:case-study-pos-maps-mean-b}, \ref{fig:case-study-pos-maps-sd-b}. Spatial patterns are observed for the values of $\widehat{a}(\xx) = E[a(\xx) \mid \yy]$ and $\widehat{b}(\xx) = E[b(\xx) \mid \yy]$ across Canada, with higher values of $\widehat{a}(\xx)$ and $\widehat{b}(\xx)$ in the Rocky mountains region in British Columbia and south of Alberta. We note that the uncertainty is higher for $b(\xx)$ in southern Nunavut and the southeast of Northwest Territories, where data is most sparse. Plotted in Figure \ref{fig:case-study-pos-maps-mean-q} and \ref{fig:case-study-pos-maps-sd-q} are the posterior $10$-year return level estimates $\widehat{z}_{10}(\xx) = E[z_{10}(\xx) \mid \yy]$ and the corresponding standard deviations. 
These two maps provide a practical guide to interpret the risk of extreme snowfall across Canada. 
For example, the posterior mean estimate of 10-year return level at the coordinates $\xx=(-80.5^{\circ}, 43.5^{\circ})$ in Waterloo, ON (marked by the red circle in Figure \ref{fig:case-study-pos-maps-mean-q} and \ref{fig:case-study-pos-maps-sd-q}) being $129.12$cm means that it takes an average of $10$ years before one observes a monthly total snowfall as extreme as $129.12$cm at this location. The 10-year return level can also be interpreted as a $10$\% chance of observing such extreme monthly total snowfalls in any year at a given location. The same claim about coordinates $\xx = (-117.5^{\circ}, 51.5^{\circ})$ in Glacier National Park in BC (marked by the red triangle) would be true for a monthly total snowfall as extreme as $377.40$ cm. Reading Figure \ref{fig:case-study-pos-maps-sd-q}, the estimate of the former location is relatively more precise compared to the latter one. From the uncertainty maps, we can also identify locations where more observations may be needed to attain a practical level of precision, e.g., at $\xx = (-130.5^{\circ}, 56.5^{\circ})$ near Granduc, BC (marked by the red diamond).

\section{Discussion}\label{sec:discussion}

In this paper we develop a computationally efficient method of Bayesian inference for fitting GEV-GP models, which have a wide range of applications in studying weather extremes. The proposed method, \lapmq, applies the Laplace approximation for calculating the marginal likelihood and a Normal approximation based on mode and quadrature
to obtain the joint posterior distribution of hyperparameters and random effects. Scalability to many thousand spatial observations is achieved using a sparsity-inducing approximation to the spatial precision matrix via GRMFs.  An efficient implementation of our method is provided in the \textsf{R}/ \textsf{C++} package \packagename. Our simulation studies indicate that the \lapmq method can produce biased estimates of the model hyperparameters.  In situations where it is important to accurately estimate these, exact Bayesian inference methods such as MCMC may be preferable.  However, when primary scientific interest lies in estimating the spatially-varying GEV parameters -- e.g., for making return-level predictions, or to perform model selection and goodness-of-fit assessments as we have done in Section~\ref{sec:case-study-select} -- then our simulations show that \lapmq is highly accurate, 
and significantly faster than a state-of-the-art MCMC algorithm on the joint posterior distribution commonly used to fit hierarchical models such as the GEV-GP.

The INLA method is a popular alternative to MCMC which has been used successfully in many analyses of weather extremes \citep[e.g.,][]{opitz18, castro-camillo19, castro-camillo20}. However, existing implementations cannot be used for GEV-GP models with two or more spatial random effects. Such flexibility can be crucial for accurate prediction of weather extreme values, as demonstrated in the analysis of extreme snowfall in Section \ref{sec:case-study}. 
Another alternative to MCMC on the full posterior distribution is the Max-and-Smooth marginal likelihood approximation, which is similar in spirit to that of Laplace, but can be much faster. However, our simulations show that it is less accurate than Laplace for estimating the shape parameter of the GEV-GP model when the number of observations per location is not sufficiently large.  


We adopted a simple Taylor approximation for the conditional posterior distribution of the random effects $p(\uu \mid \tth, \yy)$ given the fixed parameters $\tth$ of the GEV-GP model. In a sense, we have assumed that it is sufficient for estimation accuracy and precision to condition on the posterior mode of the hyperparameters. This approach might underestimate the random effects uncertainty when the posterior distribution of the hyperparameters is wide. Alternatively, a second-order Laplace approximation~\citep[e.g.,][]{tierney.etal89,bianconcini.cagnone12} can be applied to potentially increase the precision, but the cost entailed in computing the second-order term demands a trade-off between speed and precision. Numerical quadrature techniques can also be used to integrate over multiple representative values of the hyperparameters, resulting in a more precise estimation of the posterior distribution of the random effects~\cite[e.g.,][]{bianconcini14}. Applying such methods requires careful choice of the quadrature nodes and weights and other computational considerations.
\cite{stringer2021} has implemented an adaptive Gauss-Hermite quadrature technique on a flexible class of latent Gaussian models.
It is of interest to investigate its applicability to the GEV-GP model.

A fast inference method is important for the modelling and analysis of datasets containing a large number of spatial locations, especially when numerical procedures are required to work with an intractable form of the posterior distribution. Our method can work with both dense covariance matrices in the latent Gaussian processes, which scale poorly to large datasets due to the $\mathcal{O}(n^3)$ cost of covariance matrix inversions, and sparse matrices that approximate the dense covariance matrix. Indeed, computational complexity can be improved by either introducing sparsity in the covariance matrices or by reducing their dimensionality.
Methods that adopt the former approach typically involve assumptions about the dependence structure of the random effects, such as the SPDE approximation for the covariance matrix employed in this paper and in the \texttt{R-INLA} implementation. An example of the latter approach, which is another line of research, is the inducing point method within the variational inference framework \citep{quinonero05, snelson06, titsias10} which has been widely used in Gaussian process regression with tens of thousands of observations, and has also been applied on some versions of the latent Gaussian model with non-Gaussian likelihoods \citep[e.g.][]{gal-etal15, bonilla-etal19}. These methods in the Gaussian process literature are appealing from a theoretical point of view, but challenges remain in finding the optimal variational distribution that approximates the true posterior well. Extending these methods to the GEV-GP setting is a promising direction for future work.

\section*{Acknowledgements}

This work was supported by the Natural Sciences and Engineering Research Council of Canada, grant numbers RGPIN-2018-04376 (Ramezan), DGECR-2018-00349 (Ramezan) and RGPIN-2020-04364 (Lysy).

\section*{Declaration of Interest}
None.

\bibliography{mybibfile}

\newpage
\appendix
\section{Estimation Errors for the Random Effects in the Small-Scale Simulation Study}\label{app:sim-estimated-errors}
\begin{figure}[!htb]
  \centering
  \begin{subfigure}{.42\textwidth}
	\centering
      \caption{$a_\mathrm{LA}(\xx) - a(\xx)$.}\label{fig:sim-maps-diff-a}
	\includegraphics[width=\linewidth]{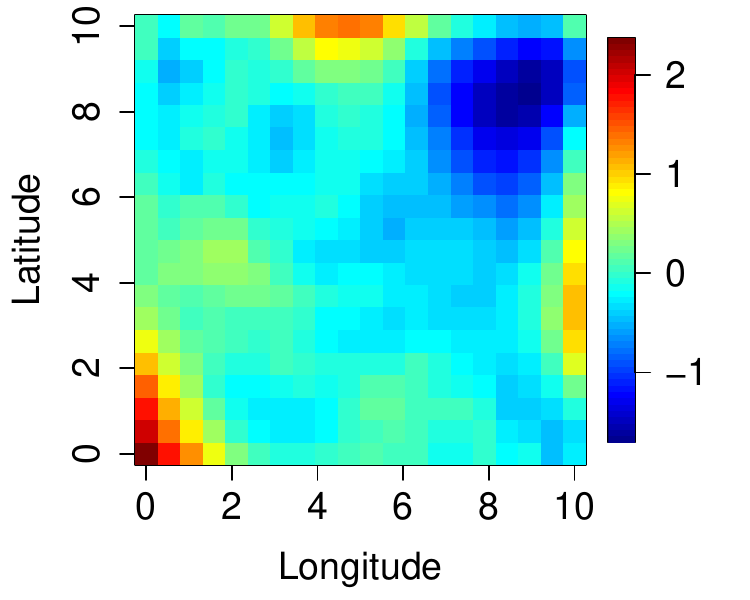}
  \end{subfigure}
  \begin{subfigure}{.42\textwidth}
	\centering
      \caption{$b_\mathrm{LA}(\xx) - b(\xx)$.}\label{fig:sim-maps-diff-b}
	\includegraphics[width=\linewidth]{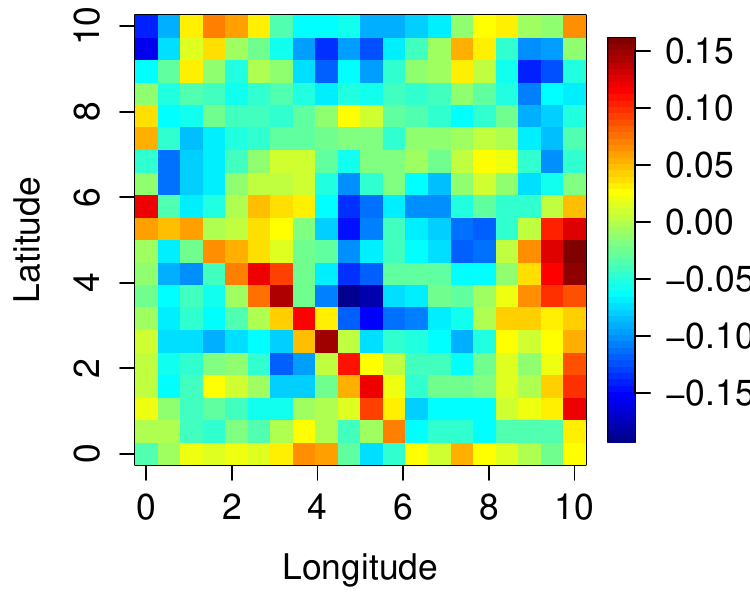}
  \end{subfigure}
  \begin{subfigure}{.42\textwidth}
	\centering
      \caption{$s_\mathrm{LA}(\xx) - s(\xx)$.}\label{fig:sim-maps-diff-s}
	\includegraphics[width=\linewidth]{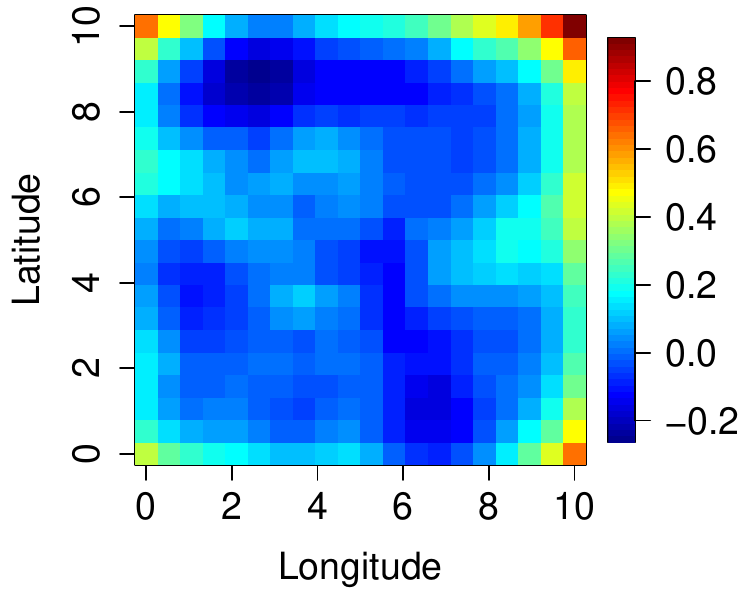}
  \end{subfigure}
  \begin{subfigure}{.42\textwidth}
	\centering
      \caption{$z_\mathrm{LA}(\xx) - z(\xx)$.}\label{fig:sim-maps-diff-z}
	\includegraphics[width=\linewidth]{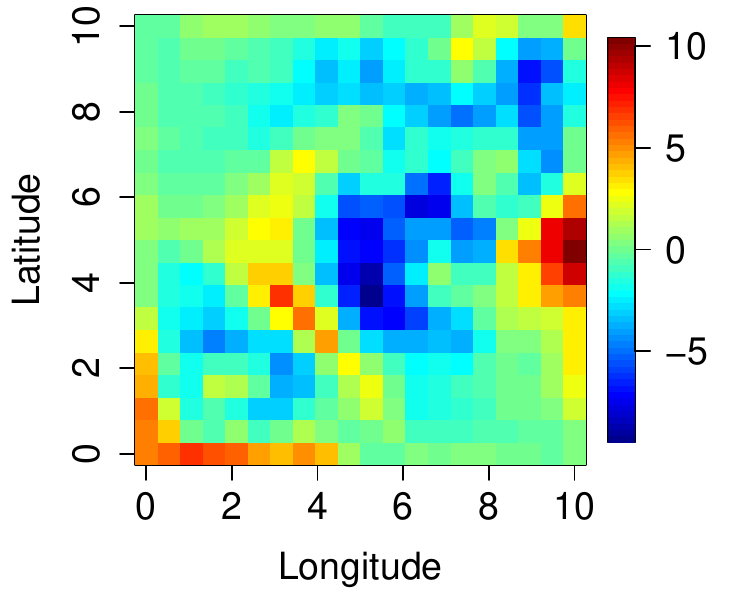}
  \end{subfigure}
  \caption{Estimation errors for $a(\xx_i)$, $b(\xx_i)$, $s(\xx_i)$ and $z_{10}(\xx_i)$ in the small-scale simulation study plotted on regular lattices.}\label{fig:sim-estimation-errors}
\end{figure}



\section{Estimation of Functions of Random Effects and Hyperparameters}\label{app:delta-method}
Consider any non-linear vector-valued function $\eeta(\uu,\tth): \mathbb{R}^{\tx{dim}(\uu)+\tx{dim}(\tth)} \to \mathbb{R}^m$. Once samples have been drawn from $\widetilde{p}(\tth\mid \yy)$ and $\widetilde{p}(\uu\mid \yy)$, which are computed using the method in Section \ref{sec:method}, samples of $p(\eeta(\uu,\tth)\mid \yy)$ are obtained by taking non-linear transformations of samples $\uu^s$ and $\tth^s$. However, when the number of locations $n$ is large, it is extremely costly to store in memory the covariance matrix of $\widetilde{p}(\uu\mid \yy)$ whose dimension is a multiple of $n$, and to sample repeatedly from a multivariate normal distribution of the same dimension. Applying the Delta method,
\begin{align}
\eeta(\uu,\tth) &\approx \eeta(\uu_{\widehat{\tth}}, \widehat{\tth}) + \JJ_{\eeta\circ\uu}\cdot (\tth-\widehat{\tth})\label{eqn:eta-delta-mean},\\
\widehat{\VV}_{\eeta}(\tth) &\approx \JJ_{\eeta} \widehat{\VV}_{\uu,\tth}(\widehat{\tth})\JJ_{\eeta}^T\label{eqn:eta-delta-v},
\end{align}
where, with $\uu_{\widehat{\tth}}=\argmin_{\uu} G(\uu; \widehat{\tth})$ and $\widehat{\VV}_{\uu}(\widehat{\tth}) = \left[\frac{\partial^2}{\partial \uu\partial \uu^T}G(\uu_{\widehat{\tth}}; \widehat{\tth})\right]^{-1}$, we have 
\begin{equation}
\begin{aligned}
    \JJ_{\eeta\circ\uu} &= \JJ_{\eeta}
     \cdot 
    \begin{bmatrix}
    \JJ_u \\
    \bm{I}
    \end{bmatrix}, 
    &
    \JJ_{\uu} = \left.\frac{\partial}{\partial \tth} \uu_{\tth} \right\vert_{\tth=\widehat{\tth}}
    & = - \widehat{\VV}_{\uu}(\widehat{\tth}) \left(\left.\frac{\partial^2}{\partial \uu \partial\tth^T}G(\uu_{\tth}, \tth)\right\vert_{\tth = \widehat{\tth}}\right)
    \\
    \JJ_{\eeta} & = \left.\begin{bmatrix}
    \frac{\partial}{\partial \uu} \eeta(\uu, \tth) & \frac{\partial}{\partial \tth} \eeta(\uu, \tth)
    \end{bmatrix}\right\vert_{\uu=\uu_{\widehat{\tth}}, \tth=\widehat{\tth}},
    & 
    \widehat{\VV}_{\uu,\tth}(\widehat{\tth}) &= \begin{bmatrix}
    \widehat{\VV}_{\uu}(\widehat{\tth}) + \JJ_{\uu}\widehat{\VV}_{\tth}\JJ_{\uu}^T & \JJ_{\uu}\widehat{\VV}_{\tth} \\
     \widehat{\VV}_{\tth}\JJ_{\uu}^T & \widehat{\VV}_{\tth}
    \end{bmatrix}.
    \end{aligned}
\end{equation}
In the special case of $\eeta(\uu,\tth)=\uu$, \eqref{eqn:eta-delta-mean} and~\eqref{eqn:eta-delta-v} are, respectively, the mean and covariance matrix of the approximate marginal posterior of $\uu$ in \eqref{eqn:normal-approx-u}.

\section{Bayesian Posterior Prediction at Arbitrary Locations}\label{app:prediction}
Let $y_\star$ denote the extreme value at 
a given spatial location $\xx_\star$, and $\uu_\star=\big(a(\xx_\star), b(\xx_\star), s(\xx_\star)\big)$ denote the corresponding random effect parameters.  Let us first assume that $\xx_\star$ is a new location, i.e., not included in the locations $\XX=(\xx_1,\ldots,\xx_n)^T$ used to fit the GEV-GP model.

Recall that $\uu=(\bm{a}, \bm{b}, \bm{s})$ is the vector of random effects at the observed locations and $\tth$ is the vector of hyperparameters. The posterior predictive distribution at the new location is given by
\begin{equation}
\begin{aligned}
p(y_\star\mid \yy) &= \int p\big(y_\star\mid a(\xx_\star), b(\xx_\star), s(\xx_\star)\big) p(\uu_\star, \tth \mid \yy) \ud \uu_\star \ud \tth\\
&= \int p(y_\star\mid a(\xx_\star), b(\xx_\star), s(\xx_\star))\left(\int p(\uu_\star \mid \uu, \tth)  p(\uu, \tth \mid \yy)\ud \uu\right)\ud \uu_\star \ud \tth\\
    &= \int p(y_\star\mid a(\xx_\star), b(\xx_\star), s(\xx_\star))\times\\
    & \ \ \ \ \ \ \ \ \ \ \ \ \left(\int p\big(a(\xx_\star)\mid\uu, \tth\big) p\big(b(\xx_\star)\mid\uu, \tth\big) p\big(s(\xx_\star)\mid\uu, \tth\big) p(\uu, \tth \mid \yy)\ud \uu\right)\ud \uu_\star \ud \tth\\
    &= \int p(y_\star\mid a(\xx_\star), b(\xx_\star), s(\xx_\star))\times\\
    & \ \ \ \ \ \ \ \ \ \ \ \  \left(\int p\big(a(\xx_\star)\mid\bm{a}, \tth\big) p\big(b(\xx_\star)\mid \bm{b}, \tth\big)p\big(s(\xx_\star)\mid\uu, \tth\big) p(\uu, \tth \mid \yy)\ud \uu\right)\ud \uu_\star \ud \tth.
\end{aligned}
\end{equation}

Note that assuming a Gaussian distribution on the location parameter $\bm{a}$, we have
\begin{equation}
    \begin{bmatrix} 
    a(\xx_\star) \\ \bm{a} \end{bmatrix} 
    \sim 
    \mathcal{N}\left(\begin{bmatrix}c(\xx_\star)'\bbe \\ c(\XX)'\bbe \end{bmatrix}, 
    \begin{bmatrix}\sigma^2_a & \boldsymbol{K}_{\tth}(\boldsymbol{x_\star}, \boldsymbol{X})\\ 
   \boldsymbol{K}_{\tth}(\boldsymbol{X}, \boldsymbol{x_\star}) & \boldsymbol{K}_{\tth}(\boldsymbol{X}, \boldsymbol{X})
    \end{bmatrix}\right).
\end{equation}
where $\XX=(\xx_1,\ldots,\xx_n)^T$, $c(\xx_\star)$ is the covariate vector at new location $\xx_\star$, $c(\XX)$ is the design matrix of the covariates at observed locations, and $[\boldsymbol{K}_{\tth}(\XX, \XX)]_{ij}=k(\xx_i, \xx_j' \mid \sigma^2, \kappa)$ using \eqref{eqn:exp-kernel}.

Hence, we can obtain $p(a(\xx_\star)\mid \uu, \tth) = p(a(\xx_\star \mid \bm{a}, \tth)$ via
\begin{equation}
a(\xx_\star)\mid \bm{a}, \tth \sim \mathcal{N}(m_{\mathrm{new}}(\boldsymbol{x_\star}), \sigma_{\mathrm{new}}(\boldsymbol{x_\star}))
\end{equation}
where  
\begin{align}
m_{\mathrm{new}}(\boldsymbol{x_\star}) &= c(\xx_\star)'\bbe+\boldsymbol{K}_{\tth}(\boldsymbol{x_\star}, \boldsymbol{X})\boldsymbol{K}_{\tth}(\boldsymbol{X}, \boldsymbol{X})^{-1}(\bm{a}-c(\XX)'\bbe),\\
\sigma_{\mathrm{new}}(\boldsymbol{x_\star})&= \sigma^2_a - \boldsymbol{K}_{\tth}(\boldsymbol{x_\star}, \boldsymbol{X})\boldsymbol{K}_{\tth}(\boldsymbol{X}, \boldsymbol{X})^{-1}\boldsymbol{K}_{\tth}(\boldsymbol{X}, \boldsymbol{x_\star}).  
\end{align}
We can obtain $p(b(\xx_\star)\mid \uu, \tth)$ and $p(s(\xx_\star)\mid \uu, \tth)$ in a similar way as above.

The posterior predictive distribution $p(y_\star\mid \yy)$ can be approximated by approximating integrals via Riemann sums and performing a multi-stage sampling scheme as follows.

1. Sample $(\bm{a}^j, \bm{b}^j, \bm{s}^j, \tth^j)$ from $p(\uu, \tth\mid \yy)$, $j=1,2,\ldots, m$.

2. For each $(\bm{a}^j, \bm{b}^j, \bm{s}^j, \tth^j)$, sample $a^j(\xx_\star)$, $b^j(\xx_\star)$ and $s^j(\xx_\star)$ from $p(a(\xx_\star)\mid \bm{a}^j, \tth^j)$, $p(b(\xx_\star)\mid \bm{b}^j, \tth^j)$ and $p(s(\xx_\star)\mid \bm{s}^j, \tth^j)$.

3. For each $\big(a^j(\xx_\star), b^j(\xx_\star), s^j(\xx_\star)\big)$, sample $y^j_\star$ from $p(y_\star\mid a^j(\xx_\star), b^j(\xx_\star), s^j(\xx_\star))$. We end up with $m$ draws of $y_\star$. 

4. The posterior mean of $y_\star$ can be approximated by $\frac{1}{m}\sum_{j=1}^{m}y^j_\star$, and any quantile of the posterior predictive distribution $p(y_\star \mid \yy)$ can be approximated by the corresponding sample quantile of $y_\star^1, \ldots, y_\star^m$.

Sampling from $p(y_\star \mid \yy)$ when $x_\star = \xx_i \in \XX$ is the $i$th observed location is similar but much easier, in that step 2 above simply consists of setting $a^j(\xx_\star)$, $b^j(\xx_\star)$, and $s^j(\xx_\star)$ to the $i$th values of $\bm{a}^j$, $\bm{b}^j$, and $\bm{s}^j$.  In either case, sampling from the posterior predictive distribution of any test statistic $p(T(y_\star) \mid \yy)$ follows immediately by setting $T^j = T(y_\star^j)$.

\end{document}